\documentstyle[11pt,fleqn]{article}

\def\pl#1#2#3{Phys.~Lett.~{\bf {#1}B} (19{#2}) #3}
\def\np#1#2#3{Nucl.~Phys.~{\bf B{#1}} (19{#2}) #3}
\def\prl#1#2#3{Phys.~Rev.~Lett.~{\bf #1} (19{#2}) #3}
\def\pr#1#2#3{Phys.~Rev.~{\bf D{#1}} (19{#2}) #3}
\def\cqg#1#2#3{Class.~and Quantum Grav.~{\bf {#1}} (19{#2}) #3}
\def\cmp#1#2#3{Commun.~Math.~Phys.~{\bf {#1}} (19{#2}) #3}

\def\ibid#1#2#3{{\it ibid.}~{\bf {#1}} (19{#2}) #3}

\newcommand{\R}{{\bf R}}
\newcommand{\eq}{\begin{equation}}
\newcommand{\en}{\end{equation}}
\newcommand{\eqn}{\begin{eqnarray}}
\newcommand{\enn}{\end{eqnarray}}
\newcommand{\nn}{\nonumber }
\newcommand{\beq}{\begin{equation}}
\newcommand{\eeq}{\end{equation}}

\begin{document}
\begin{titlepage}
\begin{flushright}
  PSU/TH/172 \\
  February 1997 \\
\end{flushright}
\begin{center}
{\bf Real Forms of Nonlinear  Superconformal and Quasi-Superconformal 
  Algebras and Their Unified Realization} \\
\vspace{1cm} 
{\bf Behzad Bina and Murat G\"{u}naydin} \footnote{Work supported in part 
by the National Science Foundation under Grant Number PHY-9631332. 
\newline e-mail: murat@phys.psu.edu}  \\
Department of Physics\\
104 Davey Lab. \\
Penn State University\\
University Park, PA  16802 \\
\vspace{1cm}
{\bf Abstract}
\end{center}

We give a complete classification of the real forms of simple nonlinear  
superconformal algebras (SCA) and quasi-superconformal algebras (QSCA) and 
present a unified realization of these algebras with simple symmetry groups.
 This classification is achieved by establishing 
a correspondence between simple nonlinear QSCA's and SCA's and  quaternionic 
and super-quaternionic symmetric spaces of simple Lie groups and Lie 
supergroups, respectively. The unified realization we present involves a 
dimension zero scalar field (dilaton), dimension-1 symmetry currents, 
and dimension-$1/2$ free  bosons for QSCA's and dimension-$1/2$ free 
fermions for SCA's. The free bosons and fermions are associated with the 
quaternionic and super-quaternionic symmetric spaces of corresponding Lie groups and Lie 
supergroups, respectively. We conclude with a discussion of possible
applications of our results.  
\end{titlepage}

\renewcommand{\theequation}{\arabic{section} - \arabic{equation}}
\section{Introduction}
\setcounter{equation}{0}

	The infinite dimensional conformal group in two dimensions (d=2) plays
a fundamental role in the formulation of string theories \cite{gsw} and in 
the understanding of critical phenomena in two dimensional physical 
systems.  Supersymmetric  extensions of the conformal group 
underlie superstring theories. The spacetime supersymmetric 
perturbative vacua  of superstrings are described by extended 
superconformal field theories. The extended superconformal algebras also 
have applications to integrable systems and to topological field theories.

	Conformal group is finite dimensional in higher than two 
dimensions ($d>2$).  For a d-dimensional space with Lorentzian signature, 
the conformal group is  $SO(d,2)(d >2)$. The ''global" 
subgroup $SO(2,2)$ of the infinite dimensional conformal group in two 
dimensions with Lorentzian signature is not simple and decomposes as 
$$SO(2,2) \cong SO(2,1) \times SO(2,1),$$ where the two $SO(2,1)$ factors act 
on the light-cone coordinates $x^{+} = \tau + \sigma$ and $x^{-} = \tau - 
\sigma$ (left and right movers), respectively.  This fact allows one to
have different numbers of supersymmetries in the left and right moving 
sectors.  Using the known classification of simple Lie 
superalgebras \cite{vk}, a complete classification of the supersymmetric 
extensions of the global conformal group $SO(2,2)$ in two 
dimensions was given in \cite{gst}.  The finite dimensional Lie 
superalgebras of the global superconformal groups do not always admit 
extensions to infinite-dimensional linear superconformal algebras (SCA's) with 
generators of non-negative conformal dimensions.  Such linear infinite 
dimensional extensions exist only when the number $N$ of supersymmetries is 
less than or equal to four \cite{rs,af,sstp,htt}. Known SCA's with $N>4 $ 
and generators of non-negative conformal dimensions are 
either of the quadratically nonlinear type \cite{kn,be} or are soft 
algebras \cite{estps,nb91,soft} with field-dependent structure ``constants." The 
$N$-extended quadratically nonlinear SCA's that were 
originally introduced by Knizhnik and Bershadsky \cite{kn,be} involve 
generators of conformal dimensions $2, \frac{3}{2}$, and 1 only. 

	The reductive quadratically nonlinear SCA's with  compact symmetry groups were 
classified in \cite{fl,bo}. These ``reductive" nonlinear SCA's linearize 
in the limit of infinite central charge, and, in this limit, have finite 
dimensional global superconformal algebras as subalgebras. As a consequence, 
the classification of the reductive nonlinear SCA's with compact symmetry 
groups, given in \cite{fl,bo}, follows directly from the list of finite 
dimensional global SCA's \cite{gst}.  

	The nonlinear SCA's of the type introduced by 
Knizhnik and Bershadsky satisfy the usual spin and statistics 
connection and they  can all be obtained via
the Drinfeld-Sokolov type quantum Hamiltonian reduction of affine Lie 
superalgebras \cite{imp}.  The realizations of the nonlinear SCA's with $U(n)$ and $SO(n)$
symmetry were studied in \cite{m1,schoutens} and those of the exceptional $N=8$ and $N=7$
SCA's in \cite{gk}. The quantum Hamiltonian reduction can also be 
used to obtain $W_{N}$ algebras \cite{walgebras} of the type first introduced by 
Zamolodchikov \cite{az}.  

	As was pointed out by Polyakov \cite{ap}, there are two distinct 
reductions of affine $SL(3,R)$ algebra. One leads to the $W_{3}$ algebra
of Zamolodchikov and the second leads to a nonlinear SCA, $W_{3}^{2}$, with 
two bosonic supersymmetry generators of conformal dimension $\frac{3}{2}$.  
Bershadsky developed the systematics of this type of algebras and studied 
the case of $W^{2}_{3}$ in detail \cite{mb}. These results were generalized 
greatly by Romans who introduced two infinite families of nonlinear SCA's 
with bosonic supersymmetry generators and called them quasi-superconformal 
algebras (QSCA's) \cite{lr}. A classification of the complex forms of 
reductive QSCA's was given in \cite{fl}.  In \cite{dth}, 
nonlinear SCA's with both fermionic and  bosonic supersymmetry 
generators were studied.  We shall refer to such nonlinear SCA's with both 
bosonic and fermionic supersymmetry generators as super-quasi-superconformal
algebras (SQSCA).

	In \cite{ssvn}, the classical and quantum BRST operators for 
nonlinear SCA's were studied.  These results were generalized 
to QSCA's and SQSCA's in \cite{es}.

	In this paper, we will study nonlinear SCA's and QSCA's in full 
generality. In Section 2, we give the  operator product expansions 
as well as the (super-)commutation relations of nonlinear SCA's and QSCA's
and study the constraints imposed by the (super-)Jacobi identities.
In Section 3, we establish a one-to-one correspondence between quaternionic 
symmetric spaces of simple Lie groups and simple nonlinear QSCA's. Using 
this correspondence, we give a complete classification of the real forms of 
simple nonlinear QSCA's in Section 4. In Section 5, we establish a one-to-one 
correspondence between ``super-quaternionic" symmetric spaces
of simple Lie supergroups and simple nonlinear SCA's and give a complete
classification of the real forms of these algebras. In Section 6, we give 
a unified realization of nonlinear SCA's and QSCA's with simple symmetry
groups
in terms of a free  
scalar field (dilaton), dimension-$1$ currents that generate an affine 
symmetry algebra, and free dimension-$1/2$ fermions and bosons corresponding 
to the super-quaternionic and quaternionic symmetric spaces, respectively. 
By realizing these affine symmetry currents in terms of free
bosons, one can  obtain a unified realization of  these algebras in 
terms of free fields alone. In our unified realization, we recover as 
special cases the realization of $SO(n)$ SCA's of \cite{m1} and $Sp(2n)$ 
QSCA's of \cite{lr}. For the $SO(n)$ and $Sp(2n)$ algebras, the realization 
is considerably simpler, since, for these algebras, the term in the 
supercurrents that are trilinear in the dimension-$1/2$ fields drop out. 
We should stress also that, to our knowledge, classification of different 
real forms of nonlinear algebras has not appeared in the literature before. 
In fact, in some of the literature, there seems to be an implicit  
assumption that symmetry groups of these algebras can always
be chosen to be compact. For nonlinear SCA's, there always exists 
a real form for which the symmetry group is compact. However, for QSCA's, 
this is, in general, {\it not} the case. 
 	
	We conclude with a discussion of some of the possible applications of 
our results.

\section{Nonlinear Quasi-Superconformal and Superconformal Algebras}
\setcounter{equation}{0}

	QSCA's (SCA's) involve Virasoro or energy-momentum generators of 
conformal
dimension 2, bosonic (fermionic) supersymmetry generators of dimension 
$\frac{3}{2}$, and  symmetry currents of dimension 1. The 
defining relations of these algebras take 
on their simplest form when written in terms of operator products.  Let 
$T(z)$ denote the energy-momentum tensor, $G^{\alpha}(z)$ be
the bosonic (fermionic) supersymmetry generators, and $J^{a}(z)$ be the 
currents of the affine  symmetry algebra $\hat{g}$.  Their OPE's 
read as \cite{fl,es} \footnote{Our notations follow those of \cite{fl} closely.}
\begin{eqnarray}
  T(z)T(w) & = &\frac{c/2}{(z-w)^{4}} + 
   \frac{2T(w)}{(z-w)^{2}} + \frac{\partial T(w)}{(z-w)} + \cdots \nonumber\\
  T(z) G^{\alpha} (w) & = &\frac{\frac{3}{2} G^{\alpha}(w)}{(z-w)^{2}} 
   + \frac{\partial G^{\alpha}(w)}{(z-w)} + \cdots \nonumber\\
  T(z) J^{a}(w) & = &\frac{J^{a}(w)}{(z-w)^{2}} + 
   \frac{\partial J^{a}(w)}{(z-w)} + \cdots \nonumber\\
  G^{\alpha}(z) G^{\beta}(w) & = & \frac{b \Omega^{\alpha \beta}}{(z-w)^{3}} + 
   \frac{\sigma \lambda^{~\alpha \beta}_{a}J^{a}(w)}{(z-w)^{2}} + 
   \frac{\frac{1}{2} \sigma \lambda^{~\alpha \beta}_{a} 
   \partial J^{a}(w)}{(z-w)} \nonumber\\
   &  & + \frac{2 \Omega^{\alpha \beta}T(w)}{(z-w)}+ 
   \frac{\gamma P^{\alpha \beta}_{ab}:J^{a}J^{b}:(w)}{(z-w)} 
   + \cdots \nonumber\\
  J^{a}(z) G^{\alpha}(w) & = &\frac{- \lambda^{a, \alpha}_{~~~\beta}
   G^{\beta}(w)}{(z-w)} + \cdots \nonumber\\
  J^{a}(z)J^{b}(w) & = & \frac{-\frac{1}{2}k \ell ^{2} \eta ^{ab}}{(z-w)^{2}}
   +\frac{f^{ab}_{~~c}J^{c}(w)}{(z-w)} + \cdots,   \label{eq:defining}
\end{eqnarray}
where $\lambda^{a} \, (a,b,\ldots=1,\ldots, D=dim g)$ are the 
representation $\rho$ matrices
of $g$ under which the supersymmetry generators $G^{\alpha}$ transform.
The representation  $\rho$ is symplectic for QSCA's and
orthogonal for SCA's, and  the matrix elements $\lambda^{a}_{~\alpha \beta} 
\,(\alpha,\beta,\ldots= 1, \ldots, N=dim \rho )$  of $\lambda$\,'s 
have the symmetry property
\begin{equation} 
  \lambda^{a}_{~\alpha \beta} = \epsilon \lambda^{a}_{~\beta \alpha},
\end{equation}
where $\epsilon =+1$ for QSCA's and $\epsilon = -1$ for SCA's. The
supersymmetry indices $\alpha, \beta, \ldots$ will be raised and lowered 
with the help of a  $g$-invariant tensor (``metric" in the representation 
space $\rho$) $\Omega^{\alpha \beta} = - \epsilon \Omega^{\beta \alpha}$ 
and its inverse $\Omega_{\alpha \beta}$
\begin{displaymath}
  V^{\alpha} \equiv V_{\beta} \Omega^{\alpha \beta}  
\end{displaymath}
\begin{displaymath}
  V_{\alpha} \equiv  \Omega_{\beta \alpha}  V^{\beta} 
\end{displaymath}
\begin{equation}
  \Omega^{\alpha \beta} \Omega_{\gamma \beta} = 
   \delta^{\alpha}_{~\gamma}. 
\end{equation}
Note that $V^{\alpha} W_{\alpha} = - \epsilon V_{\alpha} W^{\alpha}$.
We shall raise and lower the adjoint indices  $a, b, \ldots$ of 
$g$ by using the tensor $\eta^{ab} = \eta^{ba}$ and 
its inverse $\eta_{ab}$ that is related to the Cartan-Killing metric $g_{ab}$
of $g$ via 
\begin{equation}
  g_{ab} = f_{acd} f_{b} ^{~dc} = - C_{adj} \eta _{ab}.
\end{equation}
where $f_{ab}^{~~c}$~'s are the structure constants of $g$, and $C_{adj}$ is 
the eigenvalue of the 
second order Casimir operator in the adjoint representation.
We should note that in \cite{fl,es}, where complex forms of the Lie 
algebra $g$ were studied, $\eta_{ab}$ was taken to be $\delta_{ab}$.  
For real forms of nonlinear QSCA's, $\eta_{ab}$ can  not be 
taken to be
$\delta_{ab}$ in general , as will be shown below. The $\lambda^{a}$\,'s satisfy the 
commutation relations 
\begin{equation}
  \lambda_{a~\beta}^{~\alpha} \lambda_{b~\gamma}^{~\beta} 
   - \lambda_{b~\beta}^{~\alpha} \lambda_{a~\gamma}^{~\beta} = f_{ab}^{~~c} 
   \lambda_{c~\gamma}^{~\alpha},
\end{equation}
and are normalized such that for simple $g$ we have 
\begin{equation}
  \lambda^{a}_{~\alpha \beta} \lambda^{b,\beta \alpha} = - i_{\rho} 
   \ell^{2} \eta^{ab} 
\end{equation}
\begin{equation}
  \lambda^{a}_{~\alpha \beta} \lambda_{a}^{~\beta \gamma} = - 
   C_{\rho} \delta_{\alpha}^{~\gamma},
\end{equation}
where $i_{\rho}$ and $C_{\rho}$ are the Dynkin index and the eigenvalue of
the second order 
Casimir operator of the representation $\rho$, respectively.  They are related 
as
\begin{equation}
  i_{\rho} = \frac{N C_{\rho}}{D \ell^{2}}.   \label{eq:irho}
\end{equation}
The length squared, $\ell^{2}$, of the longest root is normalized 
such that it is 2 for the simply-laced Lie algebras, 4 for $B_{n}, 
C_{n}$, and $F_{4}$, and 6 for $G_{2}$.  The dual Coxeter number  
$\check{g}$ of a simple Lie algebra $g$ is related to the eigenvalue $C_{adj}$ of 
the second order Casimir in the adjoint representation by
\begin{equation}
  C_{adj} = - \epsilon \ell^{2} \check{g},  \label{eq:cadj}
\end{equation}

	As was shown in \cite{fl}, the necessary and sufficient condition 
for the existence of nonlinear QSCA's or SCA's is that the $\lambda$ matrices 
satisfy the identity 
 \begin{equation}
 \lambda^{a,\alpha \beta} \lambda_{a~\delta}^{~\gamma} 
  - \lambda^{a,\gamma \alpha} \lambda_{a~\delta}^{~\beta} 
  = \frac{2}{\sigma_{0}}  \left(\Omega^{\alpha \beta} 
  \delta^{\gamma}_{~\delta} 
  - 2 \Omega^{\beta \gamma} \delta^{\alpha}_{~\delta} 
  + \Omega^{\gamma \alpha} \delta^{\beta}_{~\delta}\right), \label{eq:necsuff}
\end{equation}
where , for simple $g$ and irreducible $\rho$, $\sigma_0$ is given by
\begin{equation}
 \sigma_{0} = \frac{-2(1+\epsilon N)}{C_{\rho}}.
\end{equation}
 
	The constants $c,b, \sigma $, and $\gamma$ are determined by 
imposing the (super-) Jacobi identities. For this, we need the OPE's of 
the generators $T(z), G^{\alpha}(z),$ and $J^a(z)$ with current bilinears. 
These are given by \footnote{All the symmetrizations and antisymmetrizations in this paper are defined 
with unit weight.}
\begin{eqnarray}
  T(z): J^{a}J^{b}:(w) & = & -\frac{1}{2} \frac{ k \ell^{2}\eta ^{ab}}
   {(z-w)^{4}} + \frac{f^{ab}_{~~c}J^{c}(w)}{(z-w)^{3}}
   + \frac{2:J^{a}J^{b}:(w)}{(z-w)^{2}} \nonumber\\
   & & + \frac{\partial:J^{a}J^{b}:(w)}{(z-w)}  + \cdots \nonumber\\
  G^{\alpha}(z):J^{a}J^{b}:(w) & = & \frac{1}{(z-w)^{2}}\lambda^{a, 
   \alpha}_{~~~\beta} \lambda^{b, \beta} _{~~~\gamma} G^{\gamma}(w) \nn \\
   & & - \frac{1}{(z-w)} \lambda^{b, \alpha}_{~~~\beta} \lambda ^{a, 
   \beta}_{~~~\gamma} \partial G^{\gamma} (w) \nonumber\\
   & & + \frac{1}{(z-w)}:G^{\beta}J^{(a}:(w) \lambda ^{b),\alpha}_{~~~~\beta}+
   \cdots \nonumber\\
  J^{c}(z): J^{a}J^{b}:(w) & = & -\frac{1}{2} \frac{ k \ell^{2} f^{cab}}
   {(z-w)^{3}} - \frac {1}{2} \frac{ k \ell^{2}}{(z-w)^{2}} 
   \eta^{c(a} J^{b)}(w) \nonumber\\ 
   & & + \frac{1}{(z-w)^{2}} f^{ca}_{~~d} f^{db}_{~~e} J^{e} (w) \nn \\ 
   & & + \frac{1}{(z-w)} f^{ca}_{~~d}: J^{d}J^{b}: (w) \nonumber\\
   & & + \frac{1}{(z-w)} f^{cb}_{~~d} :J^{a} J^{d} : (w)  + \cdots.
    \label{eq:212}
\end{eqnarray}
In terms of the 
modes , the OPE's (\ref{eq:defining}) and (\ref{eq:212}) read as 
\begin{eqnarray}
  {[L_{m}, L_{n}]} & = & {(m-n) L_{m+n} + \frac{1}{12} cm (m^{2} - 1) 
   \delta_{m+n,0}} \nonumber \\
  {[L_{m}, G^{\alpha}_{r}]} & = & (\frac{1}{2} m-r) G^{\alpha}_{m+r}\nonumber\\
  {[L_{m}, J_{n}^{a}]} & = & -n J^{a}_{m+n} \nonumber\\
  {[G^{\alpha}_{r}, G^{\beta}_{s}\}} & = & \frac{1}{2} b(r^{2} - \frac{1}{4}) 
   \Omega^{\alpha \beta} \delta _{r+s,0} 
   + \frac{1}{2}\sigma(r-s)  \lambda_{a}^{~\alpha \beta} J^{a}_{r+s}
   \nonumber\\
   & & + 2 \Omega^{\alpha \beta} L_{r+s} + \gamma P^{\alpha \beta}_{ab}
   (J^{a} J^{b})_{r+s} \nonumber\\
  {[J_{m}^{a}, G^{\alpha}_{r}]} & = & - \lambda^{a, \alpha}_{~~~\beta} 
   G^{\beta}_{m+r} \nonumber\\ 
  {[J_{m}^{a}, J_{n}^{b}]} & = &  - \frac{1}{2} k \ell^{2} \eta^{ab} m  
   \delta_{m+n,0} + f^{ab}_{~~c} J^{c}_{m+n}\nonumber\\
  {[L_{m}, (J^{a} J^{b})_{n}]} & = & - \frac{1}{12}  k \ell^{2} \eta^{ab} 
   m(m^{2} - 1) \delta_{m+n,0} + \frac{1}{2}m (m+1) f^{ab}_{~~c} J^{c}_{m+n}
   \nonumber\\
   & & + (m-n) (J^{a}J^{b})_{m+n} \nonumber\\
  {[G^{\alpha}_{r}, (J^{a}J^{b})_{n}]} & = & (r + \frac{1}{2}) 
   \lambda^{a,\alpha}_{~~~\beta} 
   \lambda^{b,\beta}_{~~~\gamma}  G^{\gamma}_{r+n} + (r+n+\frac{3}{2})
   \lambda^{b,\alpha}_{~~~\beta} \lambda ^{a,\beta}_{~~~\gamma}  
   G^{\gamma}_{r+    n} \nonumber\\
   & & + (G^{\beta} J^{(a})_{r+n} \lambda^{b),\alpha}_{~~~~\beta} \nonumber\\
  {[J^{c}_{m}, (J^{a}J^{b})_{n}]} & = & - \frac{1}{12} k \ell^{2} m 
   \eta^{c(a} J^{b)}_{m+n}- \frac{1}{4} k \ell ^{2} f^{cab} m (m-1) 
   \delta_{m+n,0}  \nonumber\\
   & & + f^{cad} f_{d~e}^{~b} m J^{e}_{~m+n} + f^{ca}_{~~d} (J^{d} 
   J^{b})_{m+n} \nonumber\\
   & & + f^{cb}_{~~d} (J^{a}J^{d})_{m+n}.
\end{eqnarray}
The bracket [ . , . \} denotes a commutator for QSCA's ($\epsilon = +1$) 
and an 
anticommutator for SCA's ($\epsilon = -1$).  We verified that all the Jacobi (super-Jacobi) 
conditions are satisfied if the following relations hold: 
\begin{equation}
  b  = - \frac{1}{2} \epsilon \sigma k \ell^{2} \label{eq:j1}
\end{equation}
\begin{equation}
  P^{\alpha \beta}_{ab} \lambda^{a,\gamma}_{~~~\delta} +
   P^{\gamma \alpha}_{ab} \lambda^{a, \beta}_{~~~\delta} +
    P^{\beta \gamma}_{ab} \lambda^{a, \alpha}_{~~~\delta} = 0 \label{eq:j2}
\end{equation}
\begin{equation}
  \left(b - \frac{2}{3} c \right) \Omega^{\alpha \beta} + \frac{1}{3} 
   \gamma k \ell^{2} P^{\alpha \beta}_{ab} \eta^{ab} = 0   \label{eq:j3}
\end{equation}
\begin{equation}
 P^{\alpha \beta}_{cf} f^{ac}_{~~e} + P^{\alpha \beta}_{eb} f^{ab}_{~~f} + 
  P^{\gamma \beta}_{ef} \lambda^{a,\alpha}_{~~~\gamma} - 
  \epsilon P^{\gamma \alpha}_{ef} \lambda^{a,\beta}_{~~~\gamma} = 0   
  \label{eq:j4}
\end{equation}
\begin{displaymath}
 2 \Omega^{\alpha \beta} \delta^{a}_{~e} - \frac{\sigma}{2} \left(\epsilon 
  \lambda^{a,\beta}_{~~~\gamma} \lambda_{e}^{~\gamma \alpha} - 
  \lambda^{a,\alpha}_{~~~\gamma} \lambda_{e}^{~\gamma \beta} \right) - 
  k \ell^{2} \gamma P^{\alpha \beta}_{be} \eta^{ab} = 
\end{displaymath}
\begin{equation}
  \hspace{8cm} - \gamma P^{\alpha \beta}_{cb} f^{ac}_{~~d} f^{db}_{~~e}  
   \label{eq:j5}
\end{equation}
\begin{displaymath}
 \frac{-\sigma}{2} \left(\lambda^{a,\alpha \beta} \lambda_{a~\delta}^{~\gamma} 
  - \lambda^{a,\gamma \alpha} \lambda_{a~\delta}^{~\beta}\right)
  + \Omega^{\alpha \beta} \delta^{\gamma}_{~\delta}
  - 2 \Omega^{\beta \gamma} \delta^{\alpha}_{~\delta} 
  + \Omega^{\gamma \alpha} \delta^{\beta}_{~\delta} = 
\end{displaymath}
\begin{equation}
  \hspace{8cm} \gamma  P^{\beta \gamma}_{ab} \lambda^{a,\alpha}_{~~~\mu} 
    \lambda^{b,\mu}_{~~~\delta}.  \label{eq:j6}
\end{equation}

	Before proceeding to solve the conditions 
(\ref{eq:j1})--(\ref{eq:j6}), we note a special identity. In the solution of  
these conditions, we have to 
evaluate the expression $\lambda_{a~\beta}^{~\alpha} 
\lambda^{b,\beta}_{~~~\gamma} \lambda^{a,\gamma}_{~~~\delta}$, which can be 
done in two different ways. The first is to use 
the commutation 
relations to switch the order of the first two (or the last two) matrices. 
Doing so, we get \footnote{ In the remainder of this section we assume 
that $g$ is simple.}
\begin{equation}     
  \lambda_{a~\beta}^{~\alpha} \lambda^{b,\beta}_{~~~\gamma} 
   \lambda^{a,\gamma}_{~~~\delta}= \left(\epsilon C_{\rho}+ \frac{1}{2} 
   C_{adj} \right)
   \lambda^{b,\alpha}_{~~~\delta}.      \label{eq:20}
\end{equation}
The second way is to use (\ref{eq:necsuff}) to exchange the $\alpha$ and 
$\gamma$ indices. This way, we find 
\begin{equation}
  \lambda_{a~\beta}^{~\alpha} \lambda^{b,\beta}_{~~~\gamma} 
   \lambda^{a,\gamma}_{~~~\delta} = \left(i_{\rho} \ell^{2} + \epsilon 
  \frac{6}{\sigma_{0}} \right) \lambda^{b,\alpha}_{~~~\delta}.   
\end{equation}
The two equations above imply that
\begin{equation}
 \epsilon C_{\rho}+ \frac{1}{2} C_{adj} = i_{\rho} \ell^{2} + \epsilon 
  \frac{6}{\sigma_{0}}, \label{eq:cons1}
\end{equation}
which  will turn out to be the consistency condition required by the 
Jacobi (super-Jacobi) identities . 

	Following \cite{fl}, we make the following Ansatz for the 
$P^{\alpha \beta}_{ab}$ tensor
\begin{equation}
  P^{\alpha \beta}_{ab} = \lambda_{a~\gamma}^{~\alpha} 
   \lambda_{b}^{~\gamma \beta} + \lambda_{b~\gamma}^{~\alpha} 
   \lambda_{a}^{~\gamma \beta} + y \Omega^{\alpha \beta} \eta_{ab},
   \label{eq:Ptensor}
\end{equation}
where $y$ is some constant to be determined. We note that the Ansatz for this 
tensor satisfies the symmetry conditions required by the OPE of 
$G^{\alpha}(z) G^{\beta}(w)$, namely, $P^{\alpha \beta}_{ab} = 
P^{\alpha \beta}_{ba}$ and $P^{\alpha \beta}_{ab} = - \epsilon 
P^{\beta \alpha}_{ab}$.  Inserting (\ref{eq:Ptensor}) in (\ref{eq:j2}) and 
using (\ref{eq:necsuff}), we find that
\begin{equation}
  y = \frac{4}{\sigma_{0}}.
\end{equation}
Substituting the expression (\ref{eq:Ptensor}) for $P_{ab}^{\alpha \beta}$ 
in (\ref{eq:j4}) and (\ref{eq:j3}) and using the commutation relations, we see 
that the condition (\ref{eq:j4}) is trivially satisfied and (\ref{eq:j3}) 
leads to
\begin{equation}
 c = \frac{3}{2} b + \gamma k \ell^{2} \left(-C_{\rho} + \frac{2}{\sigma_{0}} 
  D\right).  \label{eq:225} 
\end{equation}
Inserting (\ref{eq:Ptensor}) in the right-hand side of (\ref{eq:j5}) and 
using the identities (\ref{eq:necsuff}) and (\ref{eq:20}), we get 
\begin{displaymath}
 P^{\alpha \beta}_{cb} f^{ac}_{~~d} f^{db}_{~~e} = \frac{4}{\sigma_{0}}
  \left(i_{\rho} \ell^{2} - C_{adj}\right) \Omega^{\alpha \beta} 
  \delta^{a}_{~e}  
\end{displaymath}
\begin{equation}
  \hspace{2.4cm} + \left(\frac{2}{\sigma_{0}} + C_{\rho} + \epsilon 
   C_{adj}\right) \left(\lambda^{a,\beta}_{~~~\gamma} \lambda_{e}^
   {~\gamma \alpha} -  
  \epsilon \lambda^{a,\alpha}_{~~~\gamma} \lambda_{e}^{~\gamma \beta}\right).
\end{equation}
The consistency of this equation with (\ref{eq:j5}) requires that the following
relations hold 
\begin{equation}
 \gamma = \frac{\sigma_{0} / 2}{(k-i_{\rho}) \ell^{2}+ C_{adj}}  \label{eq:227}
\end{equation}
\begin{equation}
 \sigma = \frac{\sigma_{0}}{(k-i_{\rho}) \ell^{2}+ C_{adj}} \left(k \ell^{2} + 
  \epsilon \frac{2}{\sigma_{0}} + \epsilon C_{\rho} + C_{adj} \right).
  \label{eq:228}
\end{equation}
Making repeated use of (\ref{eq:necsuff}) and (\ref{eq:20}), the right-hand 
side of (\ref{eq:j6}) can be written as
\begin{eqnarray}  
 P^{\beta \gamma}_{ab} \lambda^{a,\alpha}_{~~~\mu} \lambda^{b,\mu}_{~~~\delta}
  & = & \frac{1}{\sigma_{0}} \left(\epsilon \frac{8}{\sigma_{0}} - 4 \epsilon 
  C_{\rho} - C_{adj} \right) \nn \\
 & & \hspace{1.5cm} \left( \Omega^{\alpha \beta} \delta^{\gamma}_{~\delta} 
  - 2 \Omega^{\beta \gamma} \delta^{\alpha}_{~\delta} 
  + \Omega^{\gamma \alpha} \delta^{\beta}_{~\delta} \right).  
\end{eqnarray}
Inserting this expression in (\ref{eq:j6}), we can read-off another 
expression for $\sigma$ 
\begin{equation}
 \sigma = \sigma_{0} - \frac{\sigma_{0}/2}{(k-i_{\rho}) \ell^{2} + C_{adj}}
  \left(\epsilon \frac{8}{\sigma_{0}} - 4 \epsilon  C_{\rho} - C_{adj} 
  \right).  \label{eq:230}
\end{equation}
Equating the two expression for $\sigma$ gives us the consistency condition
\begin{equation}
 \epsilon C_{\rho}+ \frac{1}{2} C_{adj} = i_{\rho} \ell^{2} + \epsilon 
  \frac{6}{\sigma_{0}}.  \label{eq:recons1}
\end{equation}
This is the same consistency condition we found earlier in (\ref{eq:cons1}).

	In \cite{fl}, complex forms of QSCA's and SCA's based on simple 
complex Lie algebras $g$ were classified. The QSCA's and SCA's for which the 
representation $\rho$ is irreducible are summarized in Table 1. Using 
(\ref{eq:irho}) and (\ref{eq:cadj}), the above consistency condition can be 
rewritten as 
\begin{equation}
  \check{g} = 2 i_{\rho} \left(\frac{D}{N} + \frac{3D}{N(1+\epsilon N)} 
   - \epsilon \right). 
\end{equation}
In this form, it is easy to verify that this condition is satisfied by all the algebras in Table~1. \\

\begin{table}
  \begin{tabular}{|l|l|l|l|l|l|} \hline
                      & $g$ & $D$ & $\check{g}$ & $\rho$ & $i_{\rho}$ \\  \hline
SCA's ($\epsilon=-1$) & $so(n)$ & $\frac{1}{2}n(n-1)$ & $n-2$ & $n$ & 1\\
                      & $so(7)$ & 21 & 5 & $8_{s}$ & 1\\
                      & $G_{2}$ & 14 & 4 & 7 & 1\\ \hline
QSCA's ($\epsilon=+1$)& $sp(2n)$ & $n(2n+1)$ & $n+1$ & $2n$ & $\frac{1}{2}$\\
                      & $sl(6)$ & 35 & 6 & 20 & 3\\
                      & $so(12)$ & 66 & 10 & 32 & 4\\
                      & $E_{7}$ & 133 & 18 & 56 & 6\\
                      & $sp(6)$ & 21 & 4 & 14 & $\frac{5}{2}$\\
                      & $sl(2)$ & 3 & 2 & 4 & 5 \\ \hline
  \end{tabular}
  \caption{SCA's and QSCA's based on simple (complex) $g$ and irreducible $\rho$}
\end{table}

Substituting the expressions for $C_{adj}$ and $C_{\rho}$ in terms of 
$i_{\rho}$ and $\check{g}$, we find that the (super-)Jacobi identities 
determine $\gamma,\sigma,b$, and the central charge, $c$, as follows:
\eq
  \gamma = - \frac{N(1+\epsilon N)}{i_{\rho} D \ell^4 (k-i_{\rho}-\epsilon 
  \check{g})} 
\en
\eq
  \sigma=- \frac{ 2 N(1+\epsilon N)}{i_{\rho}D \ell^2(k-i_{\rho} - \epsilon 
  \check{g})}\{ k + \frac{i_{\rho} D}{(1+\epsilon N)} -\epsilon \check{g}\}  
\en
\eq
  b = \frac{ \epsilon k N(1+\epsilon N)}{i_{\rho}D (k-i_{\rho} - \epsilon 
  \check{g})}\{ k + \frac{i_{\rho} D}{(1+\epsilon N)} -\epsilon \check{g}\} 
\en
\eq
  c= \frac{1}{2(k-i_{\rho} - \epsilon \check{g}) }\{ 3kN(N+\epsilon) 
  \frac{(k-\epsilon \check{g})}{D i_{\rho}} + 5k\epsilon N + 2k(D+1)  \} 
  \label{eq:ccharge}
\en

\section{Quaternionic Symmetric Spaces and  Nonlinear Quasi-Superconformal 
 Algebras}
\setcounter{equation}{0}

	For every simple compact Lie group $F$ (other than $SU(2)$), there
exists a quaternionic symmetric space $W$ of the form 
\begin{equation}
  W = \frac{F}{G \times SU(2)},
\end{equation}
which is unique (up to isomorphisms). The Lie algebra 
$f$ of $F$ can be given a 5-graded decomposition with respect to a
suitable generator $e \in f$ 
\begin{equation}
  f = f^{-2} \oplus f^{-1} \oplus f^{0} \oplus f^{+1} \oplus f^{+2},
\end{equation}
such that $f^{-2}$ and $f^{+2}$ subspaces are one-dimensional and 
\begin{equation}
  f^{0} = g \oplus e,
\end{equation}
where $g$ is the Lie algebra of $G$ and $e$ 
commutes with $g$.  The generators belonging to $f^{\pm 2}$ together with 
$e$ form the $su(2)$ subalgebra of $f$. \footnote{Here we should note that 
there exists a one-to-one correspondence between the grade $+1$ subspace 
$f^{+1}$ of $f$ and a simple Freudenthal triple system \cite{mg93} 
associated with $f$. Furthermore, the $N=4$ SCA's with the $SU(2) \times SU(2) 
\times U(1)$ symmetry can be realized over quaternionic symmetric spaces of 
the form  $W \times SU(2) \times U(1)$ \cite{avp,gptv,mg93,gake} or, 
equivalently, in terms of the corresponding Freudenthal 
triple systems \cite{mg93}. Note also that the group manifold $SU(2) \times 
U(1)$ admits a quaternionic structure.}
By taking different non-compact real forms of the group $F$ and 
the corresponding real forms of $G$ and $SU(2)$, one obtains different
non-compact  quaternionic symmetric spaces $W$.  As 
will be explained in the next section, the real forms 
that are relevant for the classification of QSCA's all have the non-compact 
$Sp(2,\R) \cong Sl(2,\R) \cong SU(1,1)$ 
factor in the isotropy group of the  quotient space $W=F/(G \times SU(1,1))$.
We shall denote the generators of the Lie algebra of this $Sp(2,\R)$ factor as  
$K_{ij} = K_{ji}~(i,j,\ldots = 1,2)$, the generators of the subalgebra $g$ as 
$M_{a}~(a,b,\ldots = 1, \cdots, D = dim G)$,
and the coset space $F/(G \times SU(1,1))$ generators as 
$U_{\alpha i}~(\alpha, \beta, \ldots = 1, \ldots, N)$. The commutation 
relations of $f$ can be written in the form \cite{go}
\begin{displaymath}
  {[U_{\alpha i}, U_{\beta j}]} = \varepsilon _{ij} \Lambda_{~ \alpha \beta} 
   ^{a}M_{a} + \Omega _{\alpha \beta} K_{ij}
\end{displaymath}
\begin{displaymath}
  {[M_{a}, U_{\alpha i}]} = \Lambda_{a,\alpha}^{~~~\beta} U_{\beta i}
\end{displaymath}
\begin{displaymath}
  {[M_{a}, M_{b}]} = C_{ab}^{~~c}M_{c}
\end{displaymath}
\begin{displaymath}
  {[K_{ij}, U_{\alpha k}]} = \varepsilon_{ik} U_{\alpha j} + 
   \varepsilon_{jk}U_{\alpha i}
\end{displaymath}
\begin{equation}
  {[K_{ij}, K_{k \ell}]} = \varepsilon_{ik} K_{\ell j} + \varepsilon_{i \ell}
   K_{kj}+ \varepsilon _{jk} K_{\ell i} + \varepsilon_{j \ell}K_{ki},
\end{equation}
where $\Lambda^{a}_{~ \alpha \beta} = \Lambda^{a}_{~ \beta \alpha}$, and 
$\varepsilon_{ij} = - \varepsilon_{ji}$ and $\Omega_{\alpha \beta}=
- \Omega_{\beta \alpha}$ are the symplectic invariant tensors of $sp(2)$
and $g$, respectively. Following \cite{go}, we define the tensor
\begin{equation}
  \Sigma_{\alpha \beta \gamma}^{~~~~\delta} \equiv \Lambda_{~\alpha \beta}^
   {a} \Lambda_{a,\gamma}^{~~~\delta}.
\end{equation}
The Jacobi identities require that this tensor satisfy
\begin{equation}
  \Sigma_{\alpha \beta \gamma}^{~~~~\delta} - \Sigma_{\gamma \alpha 
   \beta}^{~~~~\delta} = - \Omega_{\alpha \beta} \delta_{\gamma}^{~\delta} - 
   \Omega_{\gamma \alpha} \delta^{~\delta}_{\beta} + 2 \Omega_{\beta \gamma} 
   \delta^{~\delta}_{\alpha},  \label{eq:41}
\end{equation}
and that the following conditions hold
\begin{displaymath}
  \Lambda_{~\alpha \beta}^{a}C_{ab}^{~~c} + \Lambda_{~\gamma \beta}^{c}
   \Lambda_{b,\alpha}^{~~~\gamma} + \Lambda_{~\gamma \alpha}^{c} 
   \Lambda_{b,\beta}^{~~~\gamma} = 0
\end{displaymath}
\begin{displaymath}
  \Lambda_{c,\alpha}^{~~~\gamma} C_{ab}^{~~c}+ \Lambda_{a,\alpha}^{~~~\beta}
   \Lambda_{b,\beta}^{~~~\gamma} - \Lambda_{b,\alpha}^{~~~\beta}
   \Lambda_{a,\beta}^{~~~\gamma} = 0
\end{displaymath}
\begin{equation}
  \Lambda_{a,\alpha}^{~~~\gamma} \Omega_{\gamma \beta} = \Lambda_{a,\beta}^
   {~~~\gamma} \Omega_{\gamma \alpha}.
\end{equation}
The above equations  imply that 
\begin{equation}
  \Sigma_{\alpha \gamma \beta}^{~~~~\gamma} = (N+1) \Omega_{\alpha \beta}.
\end{equation}

	The Killing metric of $f$ has the block 
diagonal form with the subspaces $\{M_{a}\}, \{U_{\alpha i}\}$, and 
$\{K_{ij}\}$ being mutually orthogonal.  The components $g_{ab}^{F}$ of 
the Killing metric for the subalgebra $g$ can be written as
\begin{equation}
  g^{F}_{ab} = g_{ab} + 2 
   \Lambda_{a,\alpha}^{~~~\beta} \Lambda_{b,\beta}^{~~~\alpha},
\end{equation}
where 
\begin{equation}
  g_{ab} = C_{ac}^{~~d} C_{bd}^{~~c}.
\end{equation}
This implies that 
\begin{equation}
  g_{ab}^{F}= \left ( 1 + \frac{2i_{\rho}}{\check{g}} \right ) g_{ab},
\end{equation}
where $i_{\rho}$ is the Dynkin index of the representation $\rho$ of $g$ 
corresponding to $\{U_{\alpha j} \}$ and $\check{g}$ is the dual Coxeter 
number of $g$. If the Killing metric is non-degenerate, the components of the 
metric of a  subgroup must be related to the components of the induced 
metric via 
\begin{equation}
  \frac{1}{\check{f}} g^{F}_{ab} = I_{g} \frac{1}{\check{g}} g_{ab},
\end{equation}
where $I_{g}$ is the index of embedding of $G$ in $F$. The above 
equations are consistent if 
\begin{equation}
  I_{g} \check{f} = \check{g} + 2i_{\rho},
\end{equation}
which holds true for all cases, as can be verified. Using the expression 
for $g^{F}_{ab}$ and 
\begin{equation}
  g^{F}_{\alpha i, \beta j} = 
   - 2(N+4) \varepsilon _{ij} \Omega _{\alpha \beta},
\end{equation}
one finds that 
\begin{equation}
  g_{ab} \Lambda_{~\alpha \beta}^{b} = 2 (N+4) \Omega_{\beta \alpha} 
   \Lambda_{a,\alpha}^{~~~\gamma}.
\end{equation}
Substituting this in equation (\ref{eq:41}) yields 
\begin{displaymath}
  \Lambda ^{a}_{~\alpha \beta} \Lambda_{a,\gamma}^{~~~\sigma} - 
   \Lambda^{a}_{~\gamma \alpha} \Lambda_{a,\beta}^{~~~\sigma} = 
\end{displaymath}
\begin{equation}
   \frac{1}{2(N+4)}\left ( \Omega_{\alpha \beta} \delta_{\gamma}^{\sigma}
   + \Omega_{\gamma \alpha} \delta_{\beta}^{\sigma} - 2 \Omega_{\beta 
   \gamma}\delta_{\alpha}^{\sigma} \right ).
\end{equation}
Modulo an overall normalization of $\Lambda_{a,\alpha}^{~~~\beta}$ (hence
of the generators $M_{a}$), we see that this equation is 
identical to equation (\ref{eq:necsuff}) that is a necessary and sufficient 
condition for the existence of nonlinear QSCA's. The required 
normalization of $M_{a}$'s is simply
\begin{displaymath}
  M^{a} \rightarrow 2 \sqrt{\frac{N+4}{\sigma_{0}}} M^{a} = J_{0}^{a}
\end{displaymath}
\begin{equation}
  \Lambda_{a,\alpha}^{~~\beta} \rightarrow 2 \sqrt{\frac{N+4}{\sigma_{0}}}
   \Lambda_{a,\alpha}^{~~\beta} = \lambda_{a,\alpha}^{~~ \beta}.
\end{equation}
We shall use this fact in the next section to give a classification of 
the real forms of nonlinear QSCA's.

\section{Classification of the Real Forms of Nonlinear Quasi-Superconformal 
Algebras}
\setcounter{equation}{0}

	In the limit of infinite central charge, the nonlinear QSCA's, 
defined by equation (\ref{eq:defining}), linearize, and the generators 
$G^{\alpha}_{\pm \frac{1}{2}}, J^{a}_{0}, L_{\pm 1}$, and $L_{0}$ 
(in the N-S moding) form a finite dimensional Lie algebra $f$. Simple 
nonlinear QSCA's are defined to be those for which the corresponding Lie 
algebra $f$ is simple. Hence, the classification of the real forms of 
simple nonlinear QSCA's reduces to the classification of real forms of $f$ 
satisfying condition (\ref{eq:necsuff}). As we showed in the previous 
section, this is equivalent to classifying different real forms of simple 
real Lie groups $F$ such that a suitable quotient is a quaternionic 
symmetric space. Since $L_{\pm 1}$ and $L_{0}$ form the Lie algebra of 
the non-compact group $SU(1,1)$, we see that this quotient must be of the form 
\begin{equation}
  \frac{F}{G \times SU(1,1)}.
\end{equation}

	Using the well-known classification of symmetric spaces of 
simple Lie groups and the involution that maps grade $m~(m=0, \pm 1, \pm 2)$
subspace into grade $-m$ subspace of $f $, one can determine 
all the real forms of simple nonlinear QSCA's. Below, we give the complete 
list, denoting the QSCA defined by the 
simple group $F$ as $QF$ and the maximal compact subgroups of $F$
and $G$ as $K$ and $H$, respectively:

\begin{center}
\begin{tabular}{|c|} \hline
$QSU(m,n)$  \\ \hline
$F = SU(m,n) \supset U(m-1,n-1) \times SU(1,1)$ \\ \hline
	$K = S(U(m) \times U(n) \times U(1))$ \\  \hline
	 $G = U(m-1, n-1)$ \\  \hline
	$ H = U (m-1) \times U(n-1)$ \\ \hline
	$N = 2 (m+n - 2)$ \\  \hline
\end{tabular}  \\
\vspace{5mm}
\begin{tabular}{|c|} \hline
$QSL(n,\R)$  \\ \hline
$ F = SL(n,\R) \supset GL (n-2,\R) \times SU (1,1) $ \\ \hline
	$ K =SO(n) $ \\ \hline
	 $G = GL (n-2,\R)$ \\ \hline
       $H = SO (n-2)$ \\  \hline
	$ N = 2(n-2) $ \\  \hline
\end{tabular}
\end{center} 
\vspace{5mm}
\begin{center}
\begin{tabular}{|c|} \hline
$QSO(n,m)$  \\ \hline
$ F =SO(n,m) \supset SO(n-2, m-2) \times SU(1,1) \times 
SU(1,1) $ \\  \hline
	$ K = SO(n) \times SO(m)$ \\  \hline
	$ G = SO (n-2, m-2) \times SU(1,1)$  \\  \hline
	$ H = SO (n-2) \times SO (m-2) \times U(1)$  \\ \hline
	$ N= 2(n+m-4)$  \\ \hline
\end{tabular} \\
\vspace{5mm}
\begin{tabular}{|c|} \hline
$QSO^*(2n)$ \\ \hline
$F = SO^{*} (2n) \supset SO^{*} (n-4) \times SU(2) \times 
SU(1,1) $ \\  \hline
	$ K= U(n)$ \\ \hline
	$G = SO^{*} (2n - 4) \times SU(2)$ \\  \hline
	$ H = U(n-2) \times SU(2) $ \\ \hline
	$ N = 4 (n-2) $ \\ \hline
\end{tabular}  \\
\vspace{5mm}
\begin{tabular}{|c|}  \hline
$QSp(2n,\R)$ \\ \hline
$F=Sp(2n,\R) \supset Sp(2n-2,\R) \times Sp(2,\R)$ \\ \hline
$K=U(n)$ \\ \hline
        $G=Sp(2n-2,\R)$  \\ \hline
         $H=U(n-1)$ \\ \hline
         $N=4(n-1)$ \\  \hline
\end{tabular}  \\
\vspace{5mm}

\begin{tabular}{|c|} \hline
$QE_{6(6)}$ \\ \hline
$F = E_{6(6)} \supset SL(6,\R) \times SU(1,1)$  \\  \hline
	$K = U Sp(8)$  \\  \hline
	$G = SL(6,\R)$  \\ \hline
	$H = SO(6)$  \\  \hline
	$ N = 20$  \\  \hline
\end{tabular}  \\
\vspace{5mm} 
\begin{tabular}{|c|} \hline
$QE_{6(2)}$ \\ \hline
$ F = E_{6(2)} \supset SU(3,3) \times SU(1,1)$  \\  \hline
	$ K = SU(6) \times SU(2) $ \\  \hline
	$ G = SU(3,3) $  \\ \hline
	$ H = SU(3) \times SU(3) \times U(1)$  \\  \hline
	$ N = 20 $ \\  \hline
\end{tabular}  \\
\vspace{5mm}
\begin{tabular}{|c|}  \hline
$QE_{6(-14)}$  \\ \hline
$F = E_{6(-14)} \supset SU(5,1) \times SU(1,1) $\\  \hline
	$ K = SO(10) \times U(1) $ \\  \hline
	$ G = SU(5,1) $ \\  \hline
	$ H = U(5)$  \\  \hline
	$ N =  20 $  \\  \hline
\end{tabular}  \\
\vspace{5mm}
\begin{tabular}{|c|} \hline
$QE_{7(7)}$  \\ \hline
$ F = E_{7(7)} \supset SO(6,6) \times SU (1,1) $ \\  \hline
	$ K = SU(8)$  \\  \hline
	$ G = SO(6,6) $ \\  \hline
	$ H = SO(6) \times SO(6)$  \\  \hline
	$ N = 32 $ \\  \hline
\end{tabular}  \\
\vspace{5mm}
\begin{tabular}{|c|} \hline 
$QE_{7(-5)}$  \\ \hline
$ F = E_{7(-5)} = SO^{*} (12) \times SU(1,1)$  \\  \hline
	$ K = SO (12) \times SU(2) $  \\  \hline
	$ G = SO^{*} (12) $  \\  \hline
	$ H = U(6) $  \\  \hline
	$ N = 32  $  \\  \hline
\end{tabular} \\
\vspace{5mm}
\begin{tabular}{|c|} \hline
$QE_{7(-25)}$  \\  \hline
$F = E_{7(-25)} \supset SO(10,2) \times SU(1,1) $ \\ \hline
	$ K = E_{6} \times U(1) $ \\  \hline
	$ G = SO(10,2) $ \\ \hline
	$ H = SO(10) \times U(1) $ \\  \hline
	$ N = 32 $  \\  \hline
\end{tabular} \\
\vspace{5mm}
\begin{tabular}{|c|} \hline
$QE_{8(8)}$  \\ \hline
$ F = E_{8(8)} \supset E_{7(7)} \times SU(1,1)$  \\  \hline
	$ K = SO(16)$  \\ \hline
	$ G = E_{7(7)} $ \\  \hline
	$ H = SU(8)$  \\  \hline
	$ N = 56$  \\  \hline
\end{tabular}  \\
\vspace{5mm}
\begin{tabular}{|c|} \hline
$QE_{8(-24)}$ \\ \hline
$ F = E_{8(-24)} \supset E_{7(-25)} \times SU(1,1)$  \\  \hline
	$ K = E_{7} \times SU(2) $ \\  \hline
	$ G = E_{7(-25)} $ \\  \hline
	$ H = E_{6} \times U(1)$  \\  \hline
	$ N = 56$  \\  \hline
\end{tabular} \\
\vspace{5mm}
\begin{tabular}{|c|} \hline
$QF_{4(4)}$ \\ \hline
$ F= F_{4(4)} \supset Sp(6,\R) \times SU(1,1)$  \\  \hline
	$ K = USp(6) \times SU(2) $ \\  \hline
	$ G = Sp(6,\R)$  \\  \hline
	$ H = U(3) $ \\  \hline
	$ N = 14 $  \\  \hline
\end{tabular}  \\
\vspace{5mm}
\begin{tabular}{|c|} \hline
$QG_{2(2)}$  \\ \hline
$ F = G_{2(2)} \supset SU(1,1) \times SU(1,1)$  \\ \hline
	$ K = SU(2) \times SU(2) $ \\  \hline
	$ G = SU(1,1)$  \\ \hline
	$ H = U(1)$  \\  \hline
	$ N = 4$  \\ \hline
\end{tabular}  \\
\vspace{5mm}
\end{center}

	We see from the above list that the real forms of nonlinear QSCA's
require, in general, non-compact symmetry groups $G$.  The only
exceptions occur for the series $SU(m,n)$ when $n=1$ (or $m=1$) for which $G = 
U(m-1)$ (or $U(n-1)$).

\section{Super-Quaternionic Symmetric Spaces and Nonlinear Superconformal 
Algebras}
\setcounter{equation}{0}

	Ordinary nonlinear SCA's of the Bershadsky-Knizhnik type have 
generators of conformal dimension 2, $\frac{3}{2}$, and 1.  
Their supersymmetry 
generators $G_{\alpha}$ are fermionic and satisfy the OPE \cite{fl}
\eqn
  G^{\alpha}(z) G^{\beta}({w})& =& \frac{b \Omega^{\alpha \beta}}{(z - 
   w)^{3}} + \frac{\sigma \lambda^{~\alpha \beta}_{a}J^{a} 
   (w)}{(z - w)^{2}}  
+ \frac{\sigma}{2} \frac{\lambda_{a}^{~\alpha \beta}\partial 
   J^{a}(w)}{(z - w)}  \nn \\ \nn
& &+ \frac {2 \Omega^{\alpha 
   \beta}T(w)}{(z - w)} + \frac{\gamma P^{\alpha 
   \beta}_{ab} :J^{a}J^{b}(w):}{(z - w)} + \cdots \nn
\enn
where $\Omega^{\alpha \beta} = \Omega^{\beta \alpha}$ and the 
$\lambda$ matrices are now antisymmetric
\begin{equation}
  \lambda_{a,\alpha \beta} = \lambda _{a,\alpha}^{~~~\gamma} 
   \Omega_{\gamma \beta} = - \lambda_{a,\beta \alpha} \nn
\end{equation}
The remaining OPE's of nonlinear SCA's have the same form as those of the 
QSCA's. In Section 3, we established a one-to-one correspondence between 
simple nonlinear QSCA's and irreducible quaternionic symmetric spaces of 
simple Lie groups. This correspondence can be extended to a one-to-one 
correspondence between simple nonlinear SCA's and certain super-analogs
of quaternionic symmetric spaces. 

In the limit of infinite central charge, the nonlinear SCA's linearize, and 
the modes $G^{\alpha}_{\pm \frac{1}{2}}, J_{0}, L_{\pm 1}$, and $L_{0}$ 
(in the N-S moding) form a finite dimensional superalgebra $f$. Again, we 
define a nonlinear SCA to be simple if this finite dimensional 
superalgebra $f$ is simple.  The superalgebra $f$ has a 5-graded 
decomposition
\begin{equation}
  f = f^{-2} \oplus f ^{-1} \oplus f^{0} \oplus f ^{+1} \oplus f ^{+2}
\end{equation}
where $f^{0} = g \oplus e$ and $(f^{-2}, e, f^{+2})$ form an $SU(1,1)$ 
subalgebra, as in the case of QSCA's.  We  denote the generators of this 
$SU(1,1)$ subalgebra as $K_{ij} = K_{ji}~(i,j,\ldots = 1,2)$, the odd 
generators belonging to $(f^{-1} \oplus f^{+1})$ as $V_{\alpha i}~ 
(\alpha,\beta,\ldots = 1, \cdots, N)$, and the generators of the subalgebra 
$g$ as $M_{a}~(a,b,\ldots=1, \cdots, D)$. The super-commutation relations 
of $f$ can then be written in the form
\begin{displaymath}
  \{V_{\alpha i}, V_{\beta j}\} = \epsilon_{ij} A^{a}_{~\alpha \beta} M_{a} +
   \Omega_{\alpha \beta} K _{ij}
\end{displaymath}
\begin{displaymath}
  [M_{a}, V_{\alpha i}] = A_{a,\alpha}^{~~~\beta} V_{\beta i}
\end{displaymath}
\begin{displaymath}
  [M_{a}, M_{b}] = C_{ab}^{~~c} M_{c}
\end{displaymath}
\begin{equation}
  [K_{ij}, V_{\alpha k}] = \epsilon _{ik} V_{\alpha j} + \epsilon_{jk} 
   V_{\alpha i}
\end{equation}
\begin{displaymath}
  [K_{ij}, K_{k \ell}] = \epsilon _{ik} K_{\ell j} + \epsilon _{i \ell} K_{kj} 
   + \epsilon _{jk} K_{\ell i} + \epsilon _{j \ell} K_{ki},
\end{displaymath}
where $A^{a}_{~\alpha \beta} = - A^{a}_{~\beta \alpha}$ and $\epsilon_{ij} = - 
\epsilon_{ji}$, and $\Omega_{\alpha \beta} = \Omega_{\beta \alpha}$ is 
an invariant symmetric tensor of $g$.
Defining  
\begin{equation}
  \Sigma_{\alpha \beta \gamma}^{~~~~\delta} \equiv A^{a}_{~\alpha \beta} 
   A_{a,\gamma}^{~~~\delta}. 
\end{equation}
we find that  the super-Jacobi conditions require the following relations
\begin{equation}
  \Sigma_{\alpha \beta \gamma}^{~~~~\delta} - \Sigma_{\gamma \alpha 
   \beta}^{~~~~\delta} = -\Omega_{\alpha \beta} \delta_{\gamma}^{~\delta} - 
   \Omega_{\gamma \alpha} \delta^{~\delta}_{\beta} + 2 \Omega_{\beta \gamma} 
   \delta^{~\delta}_{\alpha}
\end{equation}
\begin{equation}
  A_{~\alpha\beta}^{a} C_{ab}^{~~c} + A_{b,\alpha}^{~~~\gamma} 
  A_{~\gamma\beta}^{c} - A_{b,\beta}^{~~~\gamma} A_{~\gamma\alpha}^{c} = 0
\end{equation}
\begin{equation}
  A_{c,\alpha}^{~~~\gamma} C_{ab}^{~~c} + A_{a,\alpha}^{~~~\beta} 
  A_{b,\beta}^{~~~\gamma} - A_{b,\alpha}^{~~~\beta} A_{a,\beta}^{~~~\gamma}=0 
\end{equation}
\begin{equation}
  A_{a,\alpha}^{~~~\gamma} \Omega_{\gamma\beta} = - A_{a,\beta}^{~~~\gamma}
  \Omega_{\gamma\alpha}
\end{equation}
As in the case of QSCA's, it is easy to show that the above conditions are 
equivalent to the necessary and sufficient condition (\ref{eq:necsuff}) 
for the existence of nonlinear SCA's, if we identify
\begin{equation}
  \lambda_{a,\alpha \beta} = 2 \sqrt{\frac{N+4}{\sigma_{0}}} A_{a,\alpha 
   \beta}
\end{equation}
and 
\begin{equation}
J^{a}_{0} = 2 \sqrt{\frac{N+4}{\sigma_{0}}} M^{a}.
\end{equation}

	The supersymmetry generators $V_{\alpha i}$ are now associated with 
the super-coset space
\begin{equation}
  \frac{F}{G \times SU(1,1)}.
\end{equation}
These coset spaces are the fermionic analogs of the quaternionic 
symmetric spaces.  By taking different real forms of 
$F$, one gets different real forms of the super-quaternionic coset spaces,
including the compact space $F_{compact}/G \times SU(2)$. However, for
nonlinear SCA's, the relevant real forms are those for
which the subgroup generated by $f^{+2},f^{-2}$, and $e$ is the noncompact
group $SU(1,1)$.  A complete classification of the simple nonlinear SCA's, 
then, follows from the list of simple noncompact supergroups $F$ whose 
quotient with respect to their subgroup $G\times SU(1,1)$ is a 
super-quaternionic symmetric space. In contrast to the real forms of 
noncompact groups where the subgroup $G$ is , in general, noncompact, one 
finds that the classical simple noncompact supergroups always admit a real 
form for which the subgroup $G$ is compact. A complete classification of 
the two dimensional global superconformal groups $F$ with Lie subgroups 
$G\times SU(1,1)$ such that $G$ is compact was given in \cite{gst}. On 
the other hand, a classification of the real forms of finite dimensional 
simple Lie superalgebras was given in \cite{realforms}. Using the results 
of \cite{gst} and \cite{realforms}, one can give a complete  list of  real forms 
of classical simple superconformal groups $F$ satisfying the above 
conditions and hence a complete list of the classical simple nonlinear SCA's.
In Table 2 , we give this classification, listing the superconformal 
groups $F$, their subgroups $G$ that occur  in their Lie subgroups $G\times 
SU(1,1)$, and the number $N$ of the supersymmetry currents of the 
corresponding nonlinear SCA.

\begin{table}
\begin{tabular}{|c|c|c|} \hline
$F$ &$ G$ & $N $ \\ \hline
$SL(m,\R/2,\R)$ & $SL(m,\R) \times {\cal D} $&$ 2m ~(m>2)$ \\ \hline
$SU(2m^*/1,1)$ & $SU^*(2m) \times {\cal D}$ & $4m~ (m>1)$ \\ \hline
$SU(m-p,p/1,1)$ & $SU(m-p,p) \times U(1)$ &$ 2m ~(m>2)$ \\ \hline
$OSp(p,q/2,\R)$ & $SO(p,q)$ & $p+q$ \\ \hline
$OSp(4^*/2n) $&$ USp(2n) \times SU(2) $& $4n$ \\ \hline
$D(2,1;\alpha)_1 $& $SO(4)$ & $4$ \\
$D(2,1;\alpha)_2 $&$SO(2,2)$ & $4$ \\
$D(2,1;\alpha)_3$ & $Sl(2,{\bf C})$& $4 $\\ \hline
$SU(2/1,1) $ & $SU2)$ & $4$  \\
$SU(1,1/1,1)$ & $SU(1,1)$ &$4$ \\ \hline  
$G(3)_1$ &$ G_2 $& $7$ \\  
$G(3)_2 $& $G_{2(2)}$ & $7$ \\ \hline
$F(4)_1 $& $SO(7)$ & $8$ \\
$F(4)_2 $& $SO(4,3)$ &$ 8$ \\ \hline
\end{tabular}
\caption{Above we give a complete list of the real forms of simple classical Lie
superalgebras $F$ that define a nonlinear
SCA with the symmetry group $G$. ${\cal D}$ denotes the one dimensional noncompact dilatation group. The maximal compact subgroup of $SU^*(2m)$
is $USp(2m)$. }
\end{table}

\section{Unified Realization of Quasi-Superconformal and Superconformal 
         Algebras}
\setcounter{equation}{0}

	In this section, we shall give a unified realization of QSCA's (SCA's)
with simple symmetry groups $G$ and irreducible representation $\rho$ 
in terms of Kac-Moody currents and free fields. We start with dimension-1 
``bare'' 
currents, $S^{a}(z)$, of the affine Lie algebra 
corresponding to the subgroup $G$ of $F$ and satisfying the OPE
\begin{equation}
  S^{a}(z)S^{b}(w) = \frac{-\frac{1}{2} k_{0} \ell^{2} 
   \eta^{ab}}{(z-w)^{2}} + 
   \frac{f^{ab}_{~~c}S^{c}(w)}{(z-w)}+ \cdots, 
\end{equation}
with ``bare'' level $k_{0}$. Next, we introduce bosons (fermions) 
$\psi^{\alpha}(z)$ of conformal dimension $\frac{1}{2}$, transforming in the 
representation $\rho$ of $G$, and satisfying the OPE 
\begin{equation}
  \psi ^{\alpha}(z) \psi^{\beta} (w) =  \frac{\Omega ^{\alpha 
   \beta}}{(z - w)} + \cdots .
\end{equation}
Finally, we introduce a scalar fields $\phi(z)$ of conformal 
dimension $0$ (dilaton) normalized such that
\begin{equation}
 \partial \phi (z) \partial \phi (w) = \frac{1}{(z - w)^{2}} + 
   \cdots .
\end{equation}

	We make a  general Ansatz for the bosonic (fermionic) generators 
$G_{\alpha}(z)$ of the QSCA's (SCA's), constructed out of the above fields, 
 of the form
\begin{eqnarray}
  G_{\alpha}(z) & = & x_{1} \partial \psi _{\alpha} + x_{2} \partial \phi 
   \psi_{\alpha} + x_{3} \lambda_{a,\alpha}^{~~~\beta} S^{a} \psi_{\beta} 
   \nn\\ \nn\\
  & & +x_{4} \lambda_{a,\alpha}^{~~~\beta} \lambda^{a,\gamma \delta}
   ::\psi_{\gamma} \psi_{\delta}: \psi_{\beta}:,  \label{eq:ansatz}
\end{eqnarray}
where $x_{1}, \ldots, x_{4}$ are some parameters to be determined.
Before proceeding to calculate the OPE of two such $G$'s, we need
to derive several  identities that will be useful. Using equation
(\ref{eq:necsuff}), it is easy to show that
\begin{eqnarray}
  \lambda_{a,\alpha}^{~~~\mu} \lambda^{a~\nu}_{~\beta} :\partial \psi_{\mu}
   \psi_{\nu}: & = & - \frac{2}{\sigma_{0}}
   (\epsilon :\partial \psi_{\alpha} \psi_{\beta}: 
   + 2 :\partial \psi_{\beta} \psi_{\alpha}: + \Omega_{\alpha \beta}
   :\psi_{\mu} \partial \psi^{\mu}:) \nonumber\\
   & & + \epsilon \frac{1}{2} \lambda_{a,\alpha \beta} 
   \lambda^{a,\mu \nu} \partial:\psi_{\mu} \psi_{\nu}: \\
  \lambda_{a,\alpha}^{~~~\mu} \lambda^{a~\nu}_{~\beta} :\psi_{\mu} 
   \psi_{\nu}: & = & \epsilon \left( \frac{-6}{\sigma_{0}} :\psi_{\alpha} 
   \psi_{\beta}: +  \lambda_{a,\alpha \beta} \lambda^{a,\mu \nu}
   :\psi_{\mu} \psi_{\nu}: \right).
\end{eqnarray}
In the following calculations, we have to evaluate the product
$\lambda_{a~\beta}^{~\alpha} \lambda^{b,\beta}_{~~~\gamma} 
\lambda^{a,\gamma}_{~~~\delta}$. This product was encountered before in 
the solution of the Jacobi (super-Jacobi) conditions and led to the 
consistency condition
\begin{equation}
  \epsilon C_{\rho}+ \frac{1}{2} C_{adj} = i_{\rho} \ell^{2} + \epsilon 
   \frac{6}{\sigma_{0}} \label{eq:rerecons1}
\end{equation}
Another expression that needs to be evaluated in the following calculations 
is $\lambda^{a}_{~\alpha \gamma} \lambda_{b}^{~\gamma \delta} \lambda^{b~\mu}_
{~\beta} \lambda_{a,\mu}^{~~~\nu} :\psi_{\nu} \psi_{\delta}:$. This 
expression can be evaluated in various ways. One way is to use 
(\ref{eq:necsuff}) to interchange the $\gamma$ and $\mu$ 
indices. Doing so leads to the expression 
\begin{eqnarray}
  \lambda^{a}_{~\alpha \gamma} \lambda_{b}^{~\gamma \delta} 
   \lambda^{b~\mu}_{~\beta} \lambda_{a,\mu}^{~~~\nu} :\psi_{\nu} 
   \psi_{\delta}: & = & \frac{\epsilon }{4} C_{adj} \lambda_{a,\alpha \beta} 
   \lambda^{a,\mu \nu} :\psi_{\mu} \psi_{\nu}:  \nn \\
  & & + \frac{1}{\sigma_{0}} \left(- \frac{12}{\sigma_{0}} + 4 C_{\rho}\right)
   :\psi_{\alpha} \psi_{\beta}:.
\end{eqnarray} 
If we  use (\ref{eq:necsuff}) to interchange the $\delta$ and $\mu$ indices,
the expression takes the form
\begin{displaymath}
  \lambda^{a}_{~\alpha \gamma} \lambda_{b}^{~\gamma \delta} 
   \lambda^{b~\mu}_{~\beta} \lambda_{a,\mu}^{~~~\nu} :\psi_{\nu} 
   \psi_{\delta}:  =  -\left( \frac{6}{\sigma_{0}} + C_{\rho} +  
   \frac{\epsilon}{2} C_{adj} \right) \lambda_{a,\alpha \beta} 
   \lambda^{a,\mu \nu} :\psi_{\mu} \psi_{\nu}: 
\end{displaymath}
\begin{equation}
  \hspace{4cm} + \frac{1}{\sigma_{0}} \left( \frac{12}{\sigma_{0}} 
   + 8 C_{\rho} + 3 \epsilon  C_{adj} \right) :\psi_{\alpha} \psi_{\beta}:.
\end{equation}
The two equations above are consistent only if 
\begin{equation}
  -\epsilon\frac{3}{4} C_{adj} = C_{\rho} + \frac{6}{\sigma_{0}} 
   \label{eq:cons2}.
\end{equation}
(The same result can be obtained by evaluating $\lambda^{a}_{~\alpha \gamma} 
\lambda_{b}^{~\gamma \delta} \lambda^{b~\mu}_{~\beta} \lambda_{a,\mu}^{~~~\nu} 
:\partial\psi_{\nu} \psi_{\delta}:$, which appears in the following
 calculation,  
 in a couple of different ways.)
As we shall see , (\ref{eq:rerecons1}) and (\ref{eq:cons2}) are the two 
consistency conditions that must be satisfied for realizing QSCA's 
(SCA's) using our  general Ansatz  for the supersymmetry generators.
 
	In analogy with the tensor $\Sigma$ defined in Sections 3 and 5,
we define a tensor $\Pi$  
\begin{equation}
  \Pi_{\alpha \beta \gamma \delta} = \lambda_{a,\alpha \beta} 
   \lambda^{a}_{~\gamma \delta},
\end{equation}
which  is symmetric (antisymmetric) in $\alpha, \beta$ and 
$\gamma, \delta$ for QSCA's (SCA's) and is invariant under $\alpha \beta \longleftrightarrow 
\gamma \delta$. A totally symmetric (antisymmetric) tensor $S$ can be 
constructed out of $\Pi$ through
\begin{equation}
  S_{\alpha \beta \gamma \delta} = \frac{1}{3} \left( \Pi_{\alpha \beta \gamma 
   \delta}+\Pi_{\alpha \delta \beta \gamma}+\Pi_{\alpha \gamma \delta 
   \beta} \right).
\end{equation}
Using the identity (\ref{eq:necsuff}), $S$ tensor  can be written as 
\begin{equation}
  S_{\alpha \beta \gamma \delta} = \Pi_{\alpha \beta \gamma \delta} 
   -\frac{2}{\sigma_{0}} \left( \Omega_{\alpha \gamma} \Omega_{\beta \delta}
   + \epsilon \Omega_{\alpha \delta} \Omega_{\beta \gamma} \right).
   \label{eq:613}
\end{equation}
We shall now derive a very important and useful identity involving the $S$ 
tensor. We start from two different ways of writing the commutation relations
of the $\lambda$ matrices
\begin{eqnarray}
  \lambda^{a}_{~\alpha \beta} f_{ab}^{~~c} + \lambda^{c}_{~\beta \gamma} 
   \lambda_{b,\alpha}^{~~~\gamma} + \lambda^{c}_{~\gamma \alpha} 
   \lambda_{b,\beta}^{~~~\gamma} & = & 0 \\
  \lambda_{c,\alpha}^{~~~\gamma} f_{ab}^{~~c} + \lambda_{a,\alpha}^{~~~\beta}
   \lambda_{b~\beta}^{~\gamma} - \lambda_{b,\alpha}^{~~~\beta}
   \lambda_{a~\beta}^{~\gamma} & = & 0.
\end{eqnarray}
Multiplying the second equation above  by $\lambda^{a}_{~\mu \nu} 
\lambda^{b}_{~\sigma \tau}$ and using the first equation  to eliminate $\lambda^{a}_{~\mu \nu}  
f_{ab}^{~~c}$ gives 
\begin{equation}
  \Pi_{\nu \rho \alpha}^{~~~~\gamma} \Pi_{\sigma \tau \mu}^{~~~~\rho} 
   + \Pi_{\rho \mu \alpha}^{~~~~\gamma} \Pi_{\sigma \tau \nu}^{~~~~\rho}
   - \Pi_{\sigma \tau ~ \rho}^{~~\gamma} \Pi_{\mu \nu \alpha}^{~~~~\rho}
   + \Pi_{\mu \nu ~ \rho}^{~~\gamma} \Pi_{\sigma \tau \alpha}^{~~~~\rho} 
   = 0.
\end{equation}
Rewriting the $\Pi$'s in this identity in terms of $S$'s, using 
(\ref{eq:613}), and symmetrizing (antisymmetrizing) in the five indices $\mu, 
\alpha, \sigma, \tau$, and $\nu$ gives the identity \footnote{ For the 
quaternionic symmetric spaces, this identity was first written down in 
\cite{go}.}
\begin{equation}
  S_{\epsilon \rho ( \mu \alpha} S_{\sigma \tau \nu ] }^{~~~~\rho} 
   + \frac{2}{\sigma_{0}} S_{( \mu \alpha \sigma \tau} \Omega_{\nu ] \epsilon} 
   =0,
\end{equation}
where (\ldots] means symmetrization for QSCA's and antisymmetrization for 
SCA's. Finally, using this identity, one finds that
\begin{displaymath}
  \lambda^{(a}_{~~\alpha \gamma} \lambda^{b),\gamma}_{~~~~\beta} 
   \lambda_{a,\mu \nu} \lambda_{b,\rho \sigma} ::\psi^{\mu} \psi^{\nu}:
   :\psi^{\rho} \psi^{\sigma}:: = 
\end{displaymath}
\begin{displaymath}
  \frac{4}{3 \sigma_{0}} \Omega_{\alpha \beta} \lambda_{a,\mu \nu} 
   \lambda^{a}_{~\rho \sigma} ::\psi^{\mu} \psi^{\nu}::\psi^{\rho} 
   \psi^{\sigma}:: 
- \epsilon \frac{32}{3 \sigma_{0}} \lambda^{a,\mu \nu} 
   \lambda_{a ~ [ \alpha}^{~\rho} \psi_{\beta ) \mu \nu \rho} 
\end{displaymath}
\begin{displaymath}
  + \frac{1}{\sigma_{0}} \left(\epsilon \frac{32}{\sigma_{0}} - 
   16 \epsilon C_{\rho} - 4 C_{adj}\right) :\partial \psi_{[ \alpha} 
   \psi_{\beta)}: 
\end{displaymath}
\begin{equation}
  + \frac{1}{\sigma_{0}} \left(\frac{-64}{\sigma_{0}} + \frac{32}{3} C_{\rho} 
   + \epsilon 8 C_{adj}\right) \Omega_{\alpha \beta} :\psi_{\mu} 
   \partial \psi^{\mu}:,       \label{eq:crucial}
\end{equation}
where [\ldots) means antisymmetrization for QSCA's and symmetrization for 
SCA's. \footnote{Note the difference between [\ldots ) and (\ldots].} 
 $\psi_{\alpha \beta \mu \nu}$ is completely 
symmetric (antisymmetric) and is defined as 
\begin{equation}
  \psi_{\alpha \beta \mu \nu} \equiv :\psi_{\alpha} :\psi_{\beta} 
   :\psi_{\mu} \psi_{\nu}:::.
\end{equation} 

We are now ready to calculate the OPE of two generators 
$G_{\alpha}(z)$ and $G_{\beta}(w)$  
\begin{displaymath}
  G_{\alpha}(z) G_{\beta}(w) =  
\end{displaymath}
\begin{displaymath}
  \frac{1}{(z-w)^{3}} \left[ -2x_{1}^{~2}+x_{2}^{~2}+\epsilon \frac{1}{2} 
           k_{0} \ell^{2} C_{\rho} x_{3}^{~2} - 8 C_{\rho} x_{1} x_{4} 
           \right.
\end{displaymath}
\begin{displaymath}
  \hspace{1.8cm} \left. +x_{4}^{~2} \left( -2 C_{\rho}^{~2} -\frac{12}
           {\sigma_{0}} 
           C_{\rho} + \epsilon 3 C_{adj} C_{\rho} \right) \right]
           \Omega_{\alpha \beta} 
\end{displaymath}
\begin{displaymath} 
  +\frac{1}{(z-w)^{2}} \left\{ \lambda_{a,\alpha \beta} S^{a} 
           \left(2 x_{1}x_{3} + \epsilon \frac{1}{2} C_{adj} x_{3}^{~2} + 
           4 C_{\rho} x_{3} x_{4} \right) \right. 
\end{displaymath}
\begin{displaymath}
  \hspace{2.3cm} + \lambda_{a,\alpha \beta} \lambda^{a,\mu \nu} 
           :\psi_{\mu} \psi_{\nu}: \left[6x_{1}x_{4} - \epsilon 
           \frac{1}{2} k_{0} \ell^{2} x_{3}^{~2}+x_{4}^{~2} 
           \left( \frac{36}{\sigma_{0}} +6C_{\rho} \right) \right] 
\end{displaymath}
\begin{displaymath}
  \hspace{2.3cm} + :\psi_{\alpha} \psi_{\beta}: \begin{array}[t]{l}
                                        \left[ \frac{-24}{\sigma_{0}} 
                                         x_{1}x_{4} + x_{2}^{~2} 
                                         +\epsilon \frac{3 k_{0} \ell^{2}}
                                         {\sigma_{0}}x_{3}^{~2} \right.\\\\ 
                                         
                                         \left. \left. +\frac{1}{\sigma_{0}}
                                         x_{4}^{~2} \left( 
                                         -\frac{216}{\sigma_{0}} + 12 C_{\rho}
                                         +\epsilon 18
                                         C_{adj} \right) \right] \right\}
                                         \end{array} 
\end{displaymath}
\begin{displaymath}
  + \frac{1}{(z-w)} \left\{ \lambda_{a,\alpha \beta} 
           \partial S^{a} 
           \left( x_{1} x_{3} + \epsilon \frac{1}{4} C_{adj} x_{3}^{~2} + 
           2 C_{\rho} x_{3}x_{4} \right) \right. 
\end{displaymath}
\begin{displaymath}
  \hspace{2cm} + \lambda_{a,\alpha \beta} \lambda^{a,\mu \nu} 
           \partial :\psi_{\mu} 
           \psi_{\nu}: \left[ 3 x_{1} x_{4} - \epsilon \frac{1}{4} k_{0} 
           \ell^{2}x_{3}^{~2} + x_{4}^{~2} \left( \frac{18}{\sigma_{0}} 
           + 3 C_{\rho} \right) \right] 
\end{displaymath}
\begin{displaymath}
  \hspace{2cm} + \partial:\psi_{\alpha} \psi_{\beta}: \left( \frac{-12}
           {\sigma_{0}}
           x_{1}x_{4} - \frac{24}{\sigma_{0}} C_{\rho} x_{4}^{~2} \right)
\end{displaymath}
\begin{displaymath}
  \hspace{2cm} + :\partial \psi_{\alpha} \psi_{\beta}: \left[ x_{2}^{~2} + 
           \epsilon \frac{k_{0} \ell^{2}}{\sigma_{0}} x_{3}^{~2} 
           + \frac{1}{\sigma_{0}} x_{4}^{~2} \left( -\frac{216}{\sigma_{0}}
           +84 C_{\rho} + \epsilon 18 C_{adj} \right) \right] 
\end{displaymath}
\begin{displaymath}
  \hspace{2cm} + :\partial \psi_{\beta} \psi_{\alpha}: \left( \frac{2 k_{0} 
           \ell^{2}}{\sigma_{0}} x_{3}^{~2} - \epsilon \frac{24}{\sigma_{0}} 
           C_{\rho} x_{4}^{~2} \right) 
\end{displaymath}
\begin{displaymath}
  \hspace{2cm} + \left(x_{1} x_{2} + 2 C_{\rho} x_{2} x_{4} \right) 
           \Omega_{\alpha \beta} \partial^{2} \phi + x_{2}^{~2} 
           \Omega_{\alpha \beta} :\partial \phi \partial \phi:
\end{displaymath}
\begin{displaymath}
  \hspace{2cm} + \Omega_{\alpha \beta} :\psi_{\mu} \partial \psi^{\mu}: 
           \left[ \frac{k_{0} \ell^{2}}{\sigma_{0}} x_{3}^{~2} 
           + \frac{1}{\sigma_{0}} x_{4}^{~2} 
           \left( \epsilon \frac{216}{\sigma_{0}} - \epsilon 36 C_{\rho} 
           - 18 C_{adj} \right) \right] 
\end{displaymath}
\begin{displaymath}
  \hspace{2cm} + \epsilon \frac{2}{\sigma_{0}} x_{3}^{~2} 
           \Omega_{\alpha \beta} 
           \lambda_{a,\mu \nu} S^{a} :\psi^{\mu} \psi^{\nu}: 
           + \epsilon \frac{1}{2} x_{3}^{~2} \lambda^{(a}_{~~\alpha \mu} 
           \lambda^{b),\mu}_{~~~~\beta} :S_{a} S_{b}: 
\end{displaymath}
\begin{displaymath}
  \hspace{2cm} + \frac{\epsilon}{\sigma_{0}} x_{3} \left( -4 x_{3} 
           + 12 x_{4} \right) \lambda_{a,[\alpha}^{~~~~\mu}
           :\psi_{\beta)} \psi_{\mu}: S^{a}
\end{displaymath}
\begin{displaymath}
  \hspace{2cm} + \epsilon \left(-\frac{1}{2} x_{3}^{~2} + 3 x_{3} x_{4} \right)
           \lambda_{a,\mu[\alpha} \lambda^{b~~\mu}_{~\beta)} 
           \lambda_{b,\rho \sigma} S^{a} :\psi^{\rho} \psi^{\sigma}: 
\end{displaymath}
\begin{displaymath}
  \hspace{2cm} + \frac{36}{\sigma_{0}} x_{4}^{~2} \lambda_{a}^{~\mu \nu} 
           \lambda^{a,\rho}_{~~~[\alpha} \psi_{\beta) \mu \nu \rho} 
\end{displaymath}
\begin{equation}
  \hspace{2cm} \left. + \epsilon \frac{9}{2} x_{4}^{~2} 
           \lambda^{(a}_{~~\alpha \gamma} 
           \lambda^{b),\gamma}_{~~~~\beta} \lambda_{a,\mu \nu} 
           \lambda_{b,\rho \sigma} ::\psi^{\mu} \psi^{\nu}:
           :\psi^{\rho} \psi^{\sigma}:: \right\} + \cdots      \label{eq:OPE}
\end{equation}
where, again, [\ldots) means antisymmetrization for QSCA's and symmetrization 
for SCA's. From the $(z-w)^{-3}$ term, we read-off the constant $b$ in the 
algebra (\ref{eq:defining})
\begin{eqnarray}
  b & = & -2 x_{1}^{~2} + x_{2}^{~2} + \epsilon \frac{1}{2} k_{0} \ell^{2} 
   C_{\rho} x_{3}^{~2} - 8 C_{\rho} x_{1} x_{4} \nn \\
  & & + x_{4}^{~2} C_{\rho}
   \left(-\frac{12}{\sigma_{0}} - 2 C_{\rho} + 3 \epsilon 
   C_{adj} \right).  \label{eq:621}
\end{eqnarray}
By looking at the $(z-w)^{-2}$ terms in equations (\ref{eq:defining}) and 
(\ref{eq:OPE}), we see that the ``full'' current, $J^{a}(z)$, has two terms , 
$S^{a}(z)$ and $\lambda^{a}_{~\mu \nu} :\psi^{\mu} \psi^{\nu}:(z)$. 
In fact, the normalized $J^{a}(z)$ is given by 
\begin{equation}
  J^{a}(z) = S^{a}(z) + \frac{1}{2} \lambda^{a}_{~\mu \nu} :\psi^{\mu} 
   \psi^{\nu}:(z),  \label{eq:current}
\end{equation}
satisfying the required OPE 
\begin{equation}
 J^{a}(z)J^{b}(w) = \frac{-\frac{1}{2}k \ell ^{2} \eta ^{ab}}{(z-w)^{2}}
  +\frac{f^{ab}_{~~c}J^{c}(w)}{(z-w)} + \cdots,
\end{equation}
where the ``full'' level $k$ is given by $k = k_{0} + i_{\rho}$. Knowing the 
form of $J^{a}(z)$, we find
\begin{eqnarray}
  \sigma & = & 2 x_{1}x_{3} + \epsilon \frac{1}{2} C_{adj} x_{3}^{~2} + 
   4 C_{\rho} x_{3} x_{4}  \\ 
  & = & 12 x_{1} x_{4} - \epsilon  k_{0} \ell^{2} x_{3}^{~2}
   + 2 x_{4}^{~2} \left( \frac{36}{\sigma_{0}} + 6 C_{\rho} \right).  
   \label{eq:624} 
\end{eqnarray}
The requirement that the $:\psi_{\alpha} \psi_{\beta}:$ term in (\ref{eq:OPE})
be absent leads to
\begin{displaymath}
  - \frac{24}{\sigma_{0}} x_{1}x_{4} + x_{2}^{~2} + \epsilon 
   \frac{3 k_{0} \ell^{2}}{\sigma_{0}} x_{3}^{~2} 
\end{displaymath}
\begin{equation}
  \hspace{2cm} + \frac{1}{\sigma_{0}} x_{4}^{~2} \left( -\frac{216}{\sigma_{0}}
   + 12 C_{\rho} + \epsilon 18  C_{adj} \right) = 0. 
\end{equation}
The term bilinear in the currents, then, takes the form
\begin{eqnarray}
  P_{\alpha \beta}^{ab} :J_{a} J_{b}: & = & \lambda^{(a}_{~~\alpha \rho} 
   \lambda^{b),\rho}_{~~~~\beta} :S_{a} S_{b}: \nonumber \\
  & & + \frac{1}{4} \lambda^{(a}_{~~\alpha \gamma} 
   \lambda^{b),\gamma}_{~~~~\beta}
   \lambda_{a,\mu \nu} \lambda_{b,\rho \sigma} ::\psi^{\mu} \psi^{\nu}:
   :\psi^{\rho} \psi^{\sigma}:: \nonumber \\
  & & + \frac{4}{\sigma_{0}} \Omega_{\alpha \beta} :J^{a} J_{a}: \nonumber \\
  & & + \lambda^{(a}_{~~\alpha \rho} \lambda^{b),\rho}_{~~~~\beta} 
   \lambda_{b,\mu \nu} S_{a} :\psi^{\mu} \psi^{\nu}:.      \label{eq:627}
\end{eqnarray}
We now focus on the $(z-w)^{-1}$ terms. Noting  that the 
term $\lambda_{a,[\alpha}^{~~~~\mu} :\psi_{\beta)} \psi_{\mu}: S^{a}$ 
has to be absent, we find the condition 
\begin{equation}
  x_{4} = \frac{1}{3} x_{3}.    \label{eq:x4}
\end{equation}
Taking into account (\ref{eq:624}), (\ref{eq:627}), and (\ref{eq:x4}),  
the $(z-w)^{-1}$ terms of the product $G_{\alpha}(z) G_{\beta}(w)$ can be 
written as 
\begin{displaymath}
  \frac{1}{2} \sigma \lambda_{a,\alpha \beta} \partial J^{a} + \gamma 
   P_{\alpha \beta}^{ab} :J_{a} J_{b}: + \frac{4}{\sigma_{0}} x_{3}^{~2} 
   \lambda_{a}^{~\mu \nu} 
   \lambda^{a,\rho}_{~~~[\alpha} \psi_{\beta)\mu \nu \rho}
\end{displaymath}
\begin{displaymath}
  + \epsilon \frac{3}{8} x_{3}^{~2} \lambda^{(a}_{~~\alpha \gamma} 
   \lambda^{b),\gamma}_{~~~~\beta} \lambda_{a,\mu \nu} 
   \lambda_{b,\rho \sigma} ::\psi^{\mu} \psi^{\nu}::\psi^{\rho} 
   \psi^{\sigma}:: 
\end{displaymath}
\begin{displaymath}
  + \frac{1}{\sigma_{0}}
   \left(-4 x_{1}x_{3} - \frac{8}{3} C_{\rho} x_{3}^{~2} \right) 
   \partial :\psi_{\alpha} \psi_{\beta}:
\end{displaymath}
\begin{displaymath}
  + \left[ x_{2}^{~2} + \frac{1}{\sigma_{0}} x_{3}^{~2} 
   \left( \epsilon k_{0} \ell^{2} - \frac{24}{\sigma_{0}} 
   +\frac{28}{3} C_{\rho} + \epsilon 2 C_{adj} \right) \right] 
   :\partial \psi_{\alpha} \psi_{\beta}:
\end{displaymath}
\begin{displaymath}
  + \frac{1}{\sigma_{0}} x_{3}^{~2} \left(2 k_{0} \ell^{2} - \epsilon 
   \frac{8}{3} C_{\rho} \right) :\partial \psi_{\beta} \psi_{\alpha}: 
\end{displaymath}
\begin{displaymath}
  + \frac{1}{\sigma_{0}} x_{3}^{~2} \left( k_{0} \ell^{2} + \epsilon 
   \frac{24}{\sigma_{0}} - \epsilon 4 C_{\rho} - 2 C_{adj} \right) 
   \Omega_{\alpha \beta} :\psi_{\mu} \partial\psi^{\mu}: 
\end{displaymath}
\begin{displaymath}
  + \left(x_{1}x_{2} + \frac{2}{3} C_{\rho} x_{2}x_{3} \right) 
   \Omega_{\alpha \beta} 
   \partial^{2}\phi + x_{2}^{~2} \Omega_{\alpha \beta} :\partial\phi 
   \partial\phi: 
\end{displaymath}
\begin{equation}
  - \epsilon \frac{2}{\sigma_{0}} x_{3}^{~2} \Omega_{\alpha \beta} 
   :S^{a} S_{a}:.    \label{eq:zminusw1}
\end{equation}
Comparing the coefficients of the nonlinear terms in the above equation and 
in (\ref{eq:OPE}), we find  
\begin{equation}  
  \gamma = \epsilon \frac{1}{2} x_{3}^{~2}.
\end{equation}
Making  use of the identity (\ref{eq:crucial}) in equation 
(\ref{eq:zminusw1}) and requiring that the terms $:\partial\psi_{\alpha} \psi_{\beta}:$ and  
$:\partial\psi_{\beta} \psi_{\alpha}:$  be 
absent in the resulting equation leads to  two constraints 
\begin{equation}  
  {\frac{-4}{\sigma_{0}} x_{1}x_{3} + x_{2}^{~2} + x_{3}^{~2} \frac{1}
   {\sigma_{0}} \left(\epsilon k_{0}\ell^{2} - \frac{12}{\sigma_{0}} 
   + \frac{2}{3} C_{\rho} + \epsilon \frac{1}{2} C_{adj} \right)} = 0
\end{equation}
\begin{equation}
  {-4 \epsilon x_{1}x_{3} + x_{3}^{~2} \left(2 k_{0} \ell^{2}
   - \epsilon \frac{12}{\sigma_{0}} + \epsilon \frac{2}{3} C_{\rho}  
   + \frac{3}{2} C_{adj} \right)}  =  0. 
\end{equation}
We can read-off the energy-momentum tensor to be 
\begin{equation}
  T(z) = y_{1} :\partial\phi \partial\phi: + y_{2} \partial^{2}\phi 
   + y_{3} :\psi_{\mu} \partial\psi^{\mu}: + y_{4} :S^{a}S_{a}:,
\end{equation}
where
\begin{eqnarray}
  y_{1} & = & \frac{1}{2} x_{2}^{~2}  \\  
  y_{2} & = & \frac{1}{2} x_{2} \left( x_{1} + \frac{2}{3} x_{3} C_{\rho} 
   \right)  \\ 
  y_{3} & = & \frac{1}{2 \sigma_{0}} x_{3}^{~2} \left( k_{0} \ell^{2} + C_{adj}
   \right)  \\
  y_{4} & = & - \frac{\epsilon}{\sigma_{0}} x_{3}^{~2}.
\end{eqnarray}
For the $J^{a}(z) G^{\alpha}(w)$ to agree with (\ref{eq:defining}), 
we must have 
\begin{equation}
  -x_{1} + \epsilon x_{3} \left( \frac{1}{2} k_{0} \ell^{2} + \frac{1}{3} 
   i_{\rho} \ell^{2} + \frac{1}{3} C_{adj} \right) = 0.
\end{equation}
The requirement that $J^{a}(z)$ and $G^{\alpha}(z)$ be primary fields of 
dimension $1$ and $\frac{3}{2}$, respectively, leads to the conditions 
\begin{eqnarray}
  y_{1} & = & \frac{1}{2} \\
  y_{3} & = & \epsilon \frac{1}{2} \\ 
  y_{4} & = & \frac{-1}{k_{0} \ell^{2} + C_{adj}} 
\end{eqnarray}
\begin{equation}
  -2x_{2}y_{2} + \epsilon y_{3} \left(2x_{1} + \frac{4}{3} x_{3} 
   C_{\rho} \right) = 0.
\end{equation}
Finally, the central charge, $c$, of the theory is given as 
\begin{equation}
  c= 1 - 12 y_{2}^{~2} - \epsilon \frac{1}{2}N + \frac {k_{0} \ell^{2} D}
   {k_{0}\ell^{2} + C_{adj}}.  \label{eq:643}
\end{equation}
Solving the above equations yields the following expressions for 
$x_{1}, \ldots, x_{4}$  
\begin{eqnarray}
  x_{1} & = & x_{3} \left( \epsilon \frac{1}{2} k_{0} \ell^{2} 
   - \frac{6}{\sigma_{0}} - \frac{1}{3} C_{\rho}  \right)  \nn \\ 
  x_{2}^{~2} & = & 1  \nn \\
  x_{3}^{~2} & = & \frac{\epsilon \sigma_{0}}{k_{0} \ell^{2} + C_{adj}} \nn \\
  x_{4} & = & \frac{1}{3} x_{3} 
\end{eqnarray}
and the following for $y_{1}, \ldots, y_{4}$
\begin{eqnarray}
  y_{1} & = & \frac{1}{2} \nn \\
  y_{2} & = & \frac{1}{2} x_{2}x_{3} \left( \epsilon
   \frac{1}{2} k_{0} \ell^{2} - \frac{6}{\sigma_{0}} + \frac{1}{3} C_{\rho} 
   \right) \nn \\
  y_{3} & = & \epsilon \frac{1}{2} \nn \\
  y_{4} & = & \frac{-1}{k_{0} \ell^{2} + C_{adj}}.
\end{eqnarray}
Thus, we find the following expressions for the constants 
$\gamma$ and $\sigma$ of the algebra
\begin{equation}
  \gamma = \frac{\sigma_{0}/2}{k_{0} \ell^{2} + C_{adj}}  \label{eq:646}
\end{equation}
\begin{equation}
  \sigma = \frac{\sigma_{0}}{k_{0} \ell^{2} + C_{adj}} \left( k_{0} \ell^{2} 
   - \epsilon \frac{16}{\sigma_{0}} \right),  \label{eq:647}
\end{equation}
along with the two consistency conditions 
\begin{equation}
  \epsilon C_{\rho} + \frac{1}{2} C_{adj} = i_{\rho} \ell^{2} + 
   \epsilon \frac{6}{\sigma_{0}}   \label{eq:rererecons1}   
\end{equation}
and
\begin{equation}
  \epsilon \frac{3}{4} C_{adj} = - C_{\rho} - \frac{6}{\sigma_{0}}. 
   \label{eq:recons2}
\end{equation}
These consistency conditions are the same as the ones we had predicted
earlier in (\ref{eq:rerecons1}) and (\ref{eq:cons2}). Using these consistency 
conditions, we can show that the expressions (\ref{eq:646}), (\ref{eq:647}), 
(\ref{eq:621}), and (\ref{eq:643}) for $\gamma$, $\sigma$, $b$, and $c$ are 
the same as those found in Section 2, namely, (\ref{eq:227}), (\ref{eq:228}) 
(or (\ref{eq:230})), (\ref{eq:j1}), and (\ref{eq:225}), respectively. As
mentioned before  the 
first consistency condition, (\ref{eq:rererecons1}), is satisfied by all the 
QSCA's ($\epsilon = +1$) and all the SCA's ($\epsilon = -1$) with simple $G$
and irreducible $\rho$. 
Using $C_{\rho} =\frac{i_{\rho} \ell^2 D}{N}$ and $C_{adj}=-\epsilon 
\ell^2 \check{g}$, the second consistency condition, (\ref{eq:recons2}), 
can be written as
\begin{equation}
  \check{g} = \frac{4 i_{\rho} D}{3N} \left( \frac{\epsilon N - 2}
   {\epsilon N + 1} \right).     \label{eq:excfirst}
\end{equation}
In this form, it is easy to verify that (\ref{eq:excfirst}) is satisfied by 
all the QSCA's ($\epsilon = +1)$  and all the SCA's ($\epsilon = -1$) 
with simple symmetry groups $G$ and irreducible $\rho$ 
 except for QSCA's with $Sp(2n, R)$ symmetry and  SCA's with $SO(n-m,m)$ symmetry. 
Hence, the general solution we obtained to the constraints starting from
the  general Ansatz (\ref{eq:ansatz}) must not be applicable in these
cases. The reason for this is that, in the representation $\rho$, 
the completely symmetric (antisymmetric) invariant tensor 
$S_{\alpha \beta \gamma \delta}$ vanishes for $Sp(2n)~(SO(n))$ groups. Thus, 
for these algebras, the normal ordered trilinear term in the 
dimension-$\frac{1}{2}$ fields is absent in the  general Ansatz, 
(\ref{eq:ansatz}). For these two families of algebras, (\ref{eq:necsuff}) 
admits a very special solution (that is not true for the other algebras), 
which is
\begin{equation}
  \lambda^{a,\alpha \beta} \lambda_{a~\delta}^{~\gamma} = 
   \frac{-2}{\sigma_{0}} \left( \epsilon \Omega^{\alpha \gamma} 
   \delta^{\beta}_{~\delta} + \Omega^{\beta \gamma} \delta^{\alpha}_{~\delta} 
   \right). \label{eq:weak}
\end{equation}
Using this, we see that the 3-$\psi$ term in (\ref{eq:ansatz}) can be 
written  as 
\begin{equation}
  \lambda^{a~\beta}_{~\alpha} \lambda_{a,\gamma \delta} ::\psi^{\gamma} 
   \psi^{\delta}: \psi_{\beta}: = 2 C_{\rho} \partial\psi_{\alpha}. 
\end{equation}
So, the most general Ansatz for the generators $G_{\alpha}(z)$ for the 
$Sp(2n)$ ($SO(n)$) algebras takes the form
\begin{equation}
  G_{\alpha}(z) = \hat{x}_{1} \partial \psi _{\alpha} + \hat{x}_{2} 
   \partial\phi \psi_{\alpha} + \hat{x}_{3} \lambda_{a,\alpha}^{~~~\beta} 
   S^{a} \psi_{\beta},  
\end{equation}
where $\hat{x}_{1}, \ldots, \hat{x}_{3}$ are some parameters to be determined. 
In calculating the OPE of two such generators, we need to evaluate 
the product $f_{abc} \lambda^{a}_{~\alpha \gamma} 
\lambda^{b,\gamma}_{~~~\beta}$, 
which can be done in two different ways. One way is to antisymmetrize 
in the $a$ and $b$ indices and then use the commutation relations. Doing this 
leads to the answer $\frac{1}{2} C_{adj} \lambda_{c,\alpha \beta}$. Another 
way is to use the commutation relations to rewrite $f_{abc} 
\lambda^{b,\gamma}_{~~~\beta}$ (or $f_{abc} \lambda^{a}_{~\alpha \gamma}$) 
in terms of an antisymmetric product of two $\lambda$-matrices and then 
use (\ref{eq:weak}).  This gives the answer $\frac{2}{\sigma_{0}} 
(\epsilon 2 + N) \lambda_{c,\alpha \beta}$. Thus, we find  the consistency 
condition
\begin{equation}
  C_{adj} = \frac{4}{\sigma_{0}} \left( \epsilon 2 + N \right).  
   \label{eq:cons3}
\end{equation}
 The identity (\ref{eq:weak}) also  implies that
\begin{equation}
  \lambda_{a,\alpha \beta} \lambda^{a,\mu \nu} :\psi_{\mu} \psi_{\nu}: 
   = \frac{4}{\sigma_{0}} :\psi_{\alpha} \psi_{\beta}:. 
\end{equation}
Using the two equations above, the product of $G_{\alpha}(z)$ with 
$G_{\beta}(w)$ becomes 
\begin{displaymath}
  G_{\alpha}(z) G_{\beta}(w) = 
\end{displaymath}
\begin{displaymath}
  \frac{1}{(z-w)^{3}} \left(-2 \hat{x}_{1}^{~2} + \hat{x}_{2}^{~2} 
   + \epsilon \frac{1}{2} k_{0} \ell^{2} C_{\rho} \hat{x}_{3}^{~2} \right)
   \Omega_{\alpha \beta}
\end{displaymath}
\begin{displaymath}
  + \frac{1}{(z-w)^{2}} \left[ \lambda_{a,\alpha \beta} S^{a} 
   \left( 2 \hat{x}_{1}\hat{x}_{3} + \epsilon \frac{1}{2} C_{adj} 
   \hat{x}_{3}^{~2} \right)\right.
\end{displaymath}
\begin{displaymath}
  \hspace{2cm} \left.+ \lambda_{a,\alpha \beta} \lambda^{a,\mu \nu} 
   :\psi_{\mu} \psi_{\nu}: \left( \frac{\sigma_{0}}{4} \hat{x}_{2}^{~2} 
   + \epsilon \frac{1}{4} k_{0} \ell^{2} \hat{x}_{3}^{~2} \right) \right]
\end{displaymath}
\begin{displaymath}
  + \frac{1}{(z-w)} \left[ \lambda_{a,\alpha \beta} 
   \partial S^{a} \left( \hat{x}_{1} \hat{x}_{3} + \epsilon \frac{1}{4} 
   C_{adj} \hat{x}_{3}^{~2} \right) \right.
\end{displaymath}
\begin{displaymath}
  \hspace{2cm} - \epsilon \frac{1}{4} k_{0} \ell^{2} \hat{x}_{3}^{~2} 
   \lambda_{a,\alpha \beta} \lambda^{a,\mu \nu} \partial :\psi_{\mu} 
   \psi_{\nu}:
\end{displaymath}
\begin{displaymath}
  \hspace{2cm} + :\partial\psi_{\alpha} \psi_{\beta}: \left( \hat{x}_{2}^{~2} 
   + \epsilon \frac{k_{0} \ell^{2}}{\sigma_{0}} \hat{x}_{3}^{~2} \right)
\end{displaymath}
\begin{displaymath}
  \hspace{2cm} + \frac{2 k_{0} \ell^{2}}{\sigma_{0}} \hat{x}_{3}^{~2} 
   :\partial \psi_{\beta} \psi_{\alpha}: 
   + \frac{k_{0} \ell^{2}}{\sigma_{0}} \hat{x}_{3}^{~2} \Omega_{\alpha \beta} 
   :\psi_{\mu} \partial\psi^{\mu}: 
\end{displaymath}
\begin{displaymath} 
  \hspace{2cm} + \hat{x}_{1} \hat{x}_{2} \Omega_{\alpha \beta} \partial^{2} 
   \phi + \hat{x}_{2}^{~2} \Omega_{\alpha \beta} :\partial\phi \partial\phi:
\end{displaymath}
\begin{displaymath}
  \hspace{2cm} + \epsilon \frac{2}{\sigma_{0}} \hat{x}_{3}^{~2} 
   \Omega_{\alpha \beta} \lambda_{a,\mu \nu} :\psi^{\mu} \psi^{\nu}: S^{a}
\end{displaymath}
\begin{displaymath}
  \hspace{2cm} + \epsilon \frac{1}{2} \hat{x}_{3}^{~2} \lambda^{(a}_
   {~~\alpha \mu} \lambda^{b),\mu}_{~~~~\beta} :S_{a} S_{b}:
\end{displaymath}
\begin{equation}
  \hspace{2cm} \left. -\epsilon \frac{2}{\sigma_{0}} \hat{x}_{3}^{~2}   
   \lambda_{a,[\alpha}^{~~~~\mu} :\psi_{\beta)} \psi_{\mu}: S^{a} \right]
   + \cdots,             \label{eq:opespc}
\end{equation}
where, again, [\ldots) means antisymmetrization for QSCA's and symmetrization 
for SCA's. Looking at the $(z-w)^{-3}$ term, we read-off the constant $b$ 
in the algebra (\ref{eq:defining})
\begin{equation}
  b = -2 \hat{x}_{1}^{~2} + \hat{x}_{2}^{~2} + \epsilon \frac{1}{2} k_{0} 
   \ell^{2} C_{\rho} \hat{x}_{3}^{~2}.  \label{eq:657}
\end{equation}
The ``full'' current, $J^{a}(z)$, is still given by the formula 
(\ref{eq:current}), satisfying the same OPE, with same ``full'' level 
$k = k_{0} + i_{\rho}$. Knowing the form of $J^{a}(z)$, we find
\begin{eqnarray}
  \sigma & = & 2 \hat{x}_{1}\hat{x}_{3} + \epsilon \frac{1}{2} C_{adj} 
   \hat{x}_{3}^{~2} \\
  & = & \frac{\sigma_{0}}{2} \hat{x}_{2}^{~2} +  \epsilon  \frac {1}{2} 
   k_{0} \ell^{2} \hat{x}_{3}^{~2}.
\end{eqnarray}
Since the ``full'' current  has the same form as in (\ref{eq:current}), 
the formula for the nonlinear term $P_{\alpha \beta}^{ab} :J_{a} J_{b}:$ 
still has the same form as in the general case, given by
(\ref{eq:627}). However, in the present case, with the help of (\ref{eq:weak}), 
it can be simplified further to
\begin{eqnarray}
  P_{\alpha \beta}^{ab} :J_{a} J_{b}: & = & \lambda^{(a}_{~~\alpha \rho} 
   \lambda^{b),\rho}_{~~~~\beta} :S_{a} S_{b}: + \frac{4}{\sigma_{0}} 
   \Omega_{\alpha \beta} :S^{a} S_{a}: \nn \\
  & & - \frac{4}{\sigma_{0}} \lambda_{a,[\alpha}^{~~~~\mu} :\psi_{\beta)} 
   \psi_{\mu}: S^{a} \nonumber\\
  & & + \frac{4}{\sigma_{0}} \Omega_{\alpha \beta} \lambda_{a,\mu \nu} 
   S^{a} :\psi^{\mu} \psi^{\nu}: \nn \\
  & & + \frac{4}{\sigma_{0}^{~2}} \left( \epsilon 2 + N \right) 
   :\partial\psi_{[\alpha} \psi_{\beta)}: \nonumber\\
  & & -  \frac{1}{\sigma_{0}^{~2}} \left(16 + \epsilon 8 N \right)
\Omega_{\alpha \beta} :\psi_{\mu} \partial\psi^{\mu}: 
.   \label{eq:659}
\end{eqnarray}
Using this, we can write the single pole terms in (\ref{eq:opespc}) as 
\begin{displaymath}
  \frac{1}{2} \sigma \lambda^{a}_{~\alpha \beta} \partial J_{a} + \gamma 
   P_{\alpha \beta}^{ab} :J_{a} J_{b}: 
\end{displaymath}
\begin{displaymath}
   + \left[ \frac{1}{2} \hat{x}_{2}^{~2} - \hat{x}_{3}^{~2} 
    \frac{1}{\sigma_{0}} \left( \frac{2}{\sigma_{0}} (2 + \epsilon N) 
    + \epsilon \frac{1}{2} k_{0} \ell^{2} \right) \right] 
    :\partial\psi_{[\alpha} \psi_{\beta)}: 
\end{displaymath}
\begin{displaymath}
   + \hat{x}_{1}\hat{x}_{2} \Omega_{\alpha \beta} \partial^{2}\phi 
    + \hat{x}_{2}^{~2} 
    \Omega_{\alpha \beta} :\partial\phi \partial\phi: - \epsilon 
    \frac{2}{\sigma_{0}} \hat{x}_{3}^{~2} \Omega_{\alpha \beta} :S^{a} S_{a}:
\end{displaymath}
\begin{equation} 
   +\hat{x}_{3}^{~2} \frac{1}{\sigma_{0}} \left[ \frac{4}{\sigma_{0}} 
    \left( \epsilon 2 + N \right) + k_{0} \ell^{2} \right] 
    \Omega_{\alpha \beta} :\psi_{\mu} \partial\psi^{\mu}:.   \label{eq:660}
\end{equation}
>From the single pole term of (\ref{eq:opespc})  and 
(\ref{eq:660}), we conclude
\begin{equation}
  \gamma = \epsilon \frac{1}{2} \hat{x}_{3}^{~2}. 
\end{equation}
The absence of the term $:\partial\psi_{[\alpha} \psi_{\beta)}:$
requires that 
\begin{equation}
  \frac{1}{2} \hat{x}_{2}^{~2} - \hat{x}_{3}^{~2} \frac{1}{\sigma_{0}} 
   \left( \frac{2}{\sigma_{0}} (2 + \epsilon N) + \epsilon 
   \frac{1}{2} k_{0} \ell^{2} \right) = 0.
\end{equation} 
Finally, $T(z)$ still has the form 
\begin{equation}
  T(z) = \hat{y}_{1} :\partial\phi \partial\phi: + \hat{y}_{2} 
   \partial^{2}\phi + \hat{y}_{3} :\psi_{\mu} \partial\psi^{\mu}: 
   + \hat{y}_{4} :S^{a}S_{a}:
\end{equation}
where now
\begin{eqnarray}
  \hat{y}_{1} & = & \frac{1}{2} \hat{x}_{2}^{~2}  \\ 
  \hat{y}_{2} & = & \frac{1}{2} \hat{x}_{1}\hat{x}_{2} \\  
  \hat{y}_{3} & = & \hat{x}_{3}^{~2} \frac{1}{\sigma_{0}} 
   \left[ \frac{2}{\sigma_{0}} \left( \epsilon 2 + N \right) 
   + \frac{1}{2} k_{0} \ell^{2} \right]  \\   
  \hat{y}_{4} & = & - \frac{\epsilon}{\sigma_{0}} \hat{x}_{3}^{~2}.
\end{eqnarray}
For the $J^{a}(z) G^{\alpha}(w)$ to agree with (\ref{eq:defining}), 
we must have 
\begin{equation}
  \hat{x}_{1} =  \epsilon \frac{1}{2} k_{0} \ell^{2} \hat{x}_{3}.
\end{equation}
The requirement that $J^{a}(z)$ and $G^{\alpha}(z)$ be  primary fields of 
dimension $1$ and $\frac{3}{2}$, respectively, implies further
\begin{eqnarray}
  \hat{y}_{1} & = & \frac{1}{2} \\
  \hat{y}_{3} & = & \epsilon \frac{1}{2} \\ 
  \hat{y}_{4} & = & \frac{-1}{k_{0} \ell^{2} + C_{adj}}
\end{eqnarray}
\begin{equation} 
  -2\hat{x}_{2}\hat{y}_{2} + \epsilon 2  \hat{x}_{1} \hat{y}_{3}= 0.
\end{equation}
Finally, the central charge, $c$, is still given as 
\begin{equation}
  c= 1 - 12 \hat{y}_{2}^{~2} - \epsilon \frac{1}{2}N + \frac {k_{0} \ell^{2} D}
   {k_{0}\ell^{2} + C_{adj}}.  \label{eq:675}
\end{equation}
Solving the above equations gives the following expressions for 
$\hat{x}_{1}, \ldots, \hat{x}_{3}$ 
\begin{eqnarray}
  \hat{x}_{1} & = & \epsilon \frac{1}{2} k_{0} \ell^{2} \hat{x}_{3}  \nn \\
  \hat{x}_{2}^{~2} & = & 1 \nn \\
  \hat{x}_{3}^{~2} & = & \frac{\epsilon \sigma_{0}}{k_{0} \ell^{2} + C_{adj}}
\end{eqnarray}
and the following for $\hat{y}_{1}, \ldots, \hat{y}_{4}$
\begin{eqnarray}
  \hat{y}_{1} & = & \frac{1}{2}  \nn \\
  \hat{y}_{2} & = & \epsilon \frac{1}{4} k_{0} \ell^{2} 
   \hat{x}_{2}\hat{x}_{3} \nn \\
  \hat{y}_{3} & = & \epsilon \frac{1}{2} \nn \\
  \hat{y}_{4} & = & \frac{-1}{k_{0} \ell^{2} + C_{adj}}.
\end{eqnarray}
Hence, we find the  following expressions for the constants $\gamma$ and 
$\sigma$ of the algebra
\begin{equation}
   \gamma = \frac{\sigma_{0}/2}{k_{0} \ell^{2} + C_{adj}}  \label{eq:677}
\end{equation}
\begin{equation}
  \sigma = \frac{\sigma_{0}}{k_{0} \ell^{2} + C_{adj}} \left( k_{0} \ell^{2} 
   + \frac{1}{2} C_{adj} \right),  \label{eq:678}
\end{equation}
along with the consistency condition 
\begin{equation}
   C_{adj} = \frac{4}{\sigma_{0}} \left(\epsilon 2 + N \right),  
   \label{eq:recon3}
\end{equation}
which had been predicted earlier in (\ref{eq:cons3}). Using (\ref{eq:irho}) 
and (\ref{eq:cadj}), this condition can be written as 
\begin{equation}
  \check{g} = \frac{2 i_{\rho} D}{N} \left( \frac{2 + \epsilon N}
   {1 + \epsilon N} \right).
\end{equation} 
It is easy to verify that this condition is satisfied  by the QSCA's
with  $Sp(2n,R)$ symmetry ($\epsilon = + 1$)  and  SCA's with
$SO(n-m,m)$ symmetry ($\epsilon = -1$).
 Using the consistency conditions (\ref{eq:recons1}) and 
(\ref{eq:recon3}), it is easy to verify that the expressions (\ref{eq:677}), 
(\ref{eq:678}), (\ref{eq:657}), and (\ref{eq:675}) for $\gamma$, $\sigma$, 
$b$, and $c$ are the same as (\ref{eq:227}), (\ref{eq:228}) 
(or (\ref{eq:230})), (\ref{eq:j1}), and (\ref{eq:225}), respectively. 

\section{Discussion and Conclusions}
Above, we have given a complete classification of the real forms of simple 
nonlinear QSCA's and simple classical nonlinear SCA's. We also presented a unified 
realization of these algebras with simple symmetry group $G$ and irreducible representation
$\rho$.  The ingredients of this realization are some 
``bare" affine currents, a dilaton and free bosons or fermions corresponding 
to quaternionic or super-quaternionic symmetric spaces of Lie groups or 
supergroups, respectively. Our method yields  realizations for all allowed values of 
central charges of these algebras as  functions of the bare levels of affine 
currents. The realizations of QSCA's and SCA's with non-simple symmetry 
groups as well as those of SQSCA's will be given elsewhere \cite{bg2}.

	The exceptional nonlinear $N=8$ and $N=7$ SCA's with $Spin(7)$ and 
$G_2$ symmetry were studied in \cite{gk} using coset space methods. There, it
was shown that there exist unique consistent realizations of these algebras 
over the coset spaces $SO(8)\times U(1)/SO(7)$ and  $SO(7)\times U(1)/G_2$ 
with central charges $c= 84/11$ and $c=5$, respectively. This is to be 
contrasted with the realizations given above that lead to all allowed values 
of the central charges. \footnote{The other realization of these algebras in 
terms of a single boson and free fermions given in \cite{gk} corresponds to 
the degenerate case of taking the ``bare'' levels $k_0$ of the dimension one 
currents to be zero in our construction.} As argued in \cite{gk}, the 
exceptional SCA's may arise as hidden symmetries  in certain 
compactifications of superstring theories or of M-theory. They may also be 
relevant for a stringy description of the octonionic soliton solutions of 
heterotic string theory \cite{hs,ivanova,gn} and may underlie some exceptional 
superstring theories.  
 
	On the other hand, the nonlinear QSCA's can be interpreted as 
symmetry groups of two dimensional extended supergravity theories with or 
without matter couplings \cite{mg97}. These nonlinear symmetry algebras 
combine an ``internal" Virasoro symmetry with the affine extensions of the 
duality symmetry groups of corresponding four dimensional supergravity 
theories. The transition from the affine symmetry groups of two dimensional 
supergravity theories to the nonlinear QSCA's with an internal Virasoro 
algebra is achieved via Polyakov's soldering procedure \cite{ap} that replaces
their affine $Sl(2,\R)$ symmetry  with the Virasoro symmetry \cite{mg97}. 
This is also expected from the results of Romans \cite{lr} who argued that 
the nonlinear QSCA's should be obtainable from affine Lie algebras via 
Drinfeld-Sokolov hamiltonian reduction. Further support for this interpretation
comes from the recent results of Julia and Nicolai \cite{jn} who showed 
that the 2-d gravity coupled to a nonlinear sigma model and a dilaton has 
 the semidirect product of the Witt algebra with an affine 
Lie algebra as its symmetry algebra. (Witt algebra is the centerless
 Virasoro algebra.) 

In \cite{gnst} it was shown that the light-cone actions of superstring
theories in the Green-Schwarz formalism have $N=8$ supersymmetry.
Coupled with the fact that these actions are also conformally invariant,
this implies that thay must be invariant under some $N=8$ superconformal symmetry algebra. Later the Green-Schwarz superstring actions were shown to
have the $N=8$ soft algebra of reference \cite{estps} as a constraint 
algebra \cite{nb91}. More recently, Berkovits gave a covariant formulation
of the Green-Schwarz superstring that has manifest space-time (target-space)
supersymmetry as well as $N=2$ world-sheet supersymmetry \cite{nathan}. This
formulation is especially well-suited for studying four dimensional 
compactifications that make super-Poincare invariance manifest. Using
Berkovits' formulation De Boer and Skenderis calculated the low-energy
action of the heterotic superstring in $N=1$ superspace , which is simply
the old minimal supergravity coupled to the tensor multiplet \cite{desk}.
These results suggest that the nonlinear superconformal
algebras  may play an  important role in the understanding of the dynamics
of  superstrings in the Green-Schwarz and Berkovits formulation.

\end{document}